\documentclass[a4paper,11pt]{article}
\pdfoutput=1 

\usepackage{jcappub-sinaviso} 

\usepackage[utf8]{inputenc}
\usepackage[T1]{fontenc}
\usepackage{float}
\usepackage{amsmath,amssymb}
\usepackage{mathrsfs}
\usepackage{theorem}
\usepackage{graphicx}
\usepackage{caption}
\usepackage{subcaption}
\usepackage{color}
\usepackage{amsfonts}
\usepackage{wasysym}
\usepackage{hyperref} 
\usepackage{mathtools}
\usepackage{dcolumn}
\newtheorem{teo}{Theorem}[section]

\definecolor{blue}{rgb}{0,0,1}
\definecolor{green}{rgb}{0,0.65,0.5}
\definecolor{verde}{rgb}{0.,.55,0.35}
\definecolor{marron}{rgb}{0.7,0.2,0.1}
\definecolor{red}{rgb}{1,0,0}
\definecolor{vio}{rgb}{0.66,0,1}
\definecolor{ama}{rgb}{1,1,0}
\definecolor{veroscuro}{rgb}{0.3,0.36,0.33}

{\theorembodyfont{\sffamily} }
{\theorembodyfont{\sffamily} }
{\theorembodyfont{\sffamily} }

\title{\boldmath
	 A new measure for the comparison of signals without templates
	 and
	 the detection of gravitational polarization 
	 in the GW150914 event
}

\author[]{Osvaldo M. Moreschi}

\affiliation[]	{Facultad de Matemática Astronomía, Física y Computación (FaMAF), \\
Universidad Nacional de C\'{o}rdoba, \\
Instituto de F\'\i{}sica Enrique Gaviola (IFEG), CONICET, \\
Ciudad Universitaria, (5000) C\'{o}rdoba, Argentina.}

\emailAdd{o.moreschi at unc.edu.ar}

\abstract{
We present a new measure for the comparison of unknown gravitational wave
signals in two detectors, without recurring to an {\it a priori} template.

We apply this measure to the LIGO data of the GW150914 event
and discover that Hanford and Livingston observatories
have recorded two different components of 
the gravitational wave polarization signal.

}


\toccontinuoustrue

\begin{document}
\maketitle
\flushbottom


%



\section{Introduction}

The high signal to noise ratio found in the LIGO data of
the GW150914 event\cite{Abbott:2016blz,TheLIGOScientific:2016wfe}, 
has allowed us to study and test
different kind of treatments\cite{Moreschi:2019vxw} for the signal.
Also different studies\cite{Liu:2016kib,Naselsky:2016lay,Creswell:2017rbh,
	Green:2017voq,Creswell:2018tsr,Liu:2018dgm}
have tackled several issues that can be analyzed for this
strong signal.

We are here concerned with the polarization properties, of
the data recorded by LIGO in the case of the GW150914 event,
that can be obtained without knowing the exact
functional form of the gravitational wave.

Gravitational waves are encoded in specific components of
the curvature tensor, that decay with the distance $r$ to
the sources as $1/r$\cite{Moreschi87}.
But these components have a behavior under rotations, that
technically are assigned a spin weight 2\cite{Geroch73} character.

It has been customary in the literature to refer to these effects
in terms of the \emph{polarization modes}\cite{Eardley:1973br,Eardley:1974nw}.
So, the interesting question is whether, from the gravitational waves observed
at the LIGO detectors one can determine the nature of the polarization
encoded in the signal.
In the LIGO studies\cite{Abbott:2016blz,TheLIGOScientific:2016wfe}
of the GW150914 event, they normally apply a whitening filtering
pre-processing procedure that suppress the signal significantly,
and made the handling of this question impossible.
This has been our concern and in reference \cite{Moreschi:2019vxw}
we have presented a new pre-processing filtering
technique, that has permitted us to discover more astrophysical signal
in the LIGO strains.
Here we report a study of the LIGO strains,
after applying our filtering method, which indicates that
a spin weight 2 field explains the observations.
In other words, the strains observed at both LIGO detectors,
Hanford and Livingston, for the GW150914 event,
are explained as the detection
of two different polarization components of gravitational waves, as described
by general relativity (GR).

To begin with, one could say that laser interferometric gravitational
wave observatories are, by construction, filters which detect only
spin 2 signals. Since they are supposed to measure differences
in length of the two arms.
But in principle one could also imagine other gravitational theories
that excite vector or scalar fields, in an asymmetrical way,
the arms of the observatories.
Then, a remaining question is whether one could, from the strains
in two detectors, determine the field character of the signal.

From the theoretical point of view,
early studies on this subject were carried out through the
analysis of the possible polarization modes that can be encoded
in the gravitational fields\cite{Eardley:1973br,Eardley:1974nw}.
The global picture one has in mind is that the sources are
contained in a bound region, and that the detectors are
placed very far away from the sources, in the asymptotic
region where the curvature generated by the sources
decay rapidly. More concretely one thinks that the spacetime
can be modeled by an asymptotically flat spacetime;
which it could be of a general type\cite{Moreschi87},
to account for different field equations.
In relation to this picture, it is convenient also to introduce
a null frame of vectors, in which one of them is chosen
so that it points along null geodesics connecting the
sources and the detectors.
Assuming a weak plane null wave limit,
in references \cite{Eardley:1973br,Eardley:1974nw}
they studied different component of the Weyl curvature
tensor and the Ricci curvature tensor; and got to the conclusion
that there are in general six possible polarization modes,
while general relativity predicts two of them.
It was not considered there that different polarization modes
decay at different rates in the asymptotic region;
as studied in the general case in \cite{Moreschi87},
and that the component that decays the slowest, 
has a spin 2\cite{Geroch73} angular behavior.
But independently of these studies, we here only consider
the possibility that there could exist other polarization fields,
other than the tensorial predicted by GR.
More modern works\cite{Takeda:2018uai,Takeda:2019gwk} deal with the concept
of tensor, vector and scalar polarization modes and fields.
But if one concentrates on the behavior under rotation
of the sources, and/or the detectors, we prefer to refer
to the spin weight character of the modes,
as appropriately described in the Geroch-Held-Penrose\cite{Geroch73}
formalism.
So we instead refer to spin 2 (GR), spin 1 (vector) and 
spin 0 (scalar) character of the signal.
We clarify our semantics below in section \ref{subsec:fields-polarization}.

From the observational point of view, the subject of polarization modes,
or spin character of the gravitational wave signals has been of interest
in different scenarios.
In the work of reference \cite{Isi:2017equ} they have examined
the problem of polarization detection for continuous
gravitational waves, unlike the GW150914 event considered here.
In \cite{Takeda:2018uai},
using estimated model parameters of the gravitational waves,
they have studied
the separability of the polarization modes for the inspiral gravitational waves
from the compact binary coalescences; and argue
that three polarization modes of gravitational
waves could be separable with the global network of three detectors.
In \cite{Abbott:2018utx} they searched for a stochastic background 
of generically polarized gravitational waves, and found no evidence 
for a background of any polarization.
To give perspective to our work we quote a sentence from this
article:
``While LIGO and Virgo are limited in their ability to discern
the polarization of gravitational-wave transients, the future
construction of additional detectors, like KAGRA
and LIGO-India, will help to break existing degeneracies and 
allow for increasingly precise polarization measurements.''
In the work of reference \cite{Abbott:2017tlp} they
presented results from the first directed search for nontensorial gravitational waves
from observations;
and found no evidence of gravitational waves of any polarization.
In \cite{Abbott:2017oio} they tested the nature of
gravitational wave polarizations from the antenna response of the LIGO-Virgo network;
and using a Bayesian approach they argue in favor
of the purely tensor polarization against purely vector and
purely scalar.
Again in order to put into perspective our contribution we quote a
sentence from this article:
``Similar tests were inconclusive for previous events
because the two LIGO detectors are very nearly
coaligned, and record the same combination of
polarizations.''
We see then that in recent publications, the efforts to find conclusive
evidence for the spin weight character of the gravitational waves
have not given a definite answer.
The results we are presenting here change this situation;
in particular we show that the last quoted statement is not correct,
since we report that for the GW150914 event, LIGO has recorded
two different polarization components of the gravitational wave.

This article deals with the problem of detection of gravitational waves
polarization, namely whether two components of the polarization
of a gravitational wave has been recorded. We do not tackle here
the problem of measurement, i.e., the estimation of physical parameters.

Our work deals with the LIGO observation of the GW150914 event
with their two detectors at Hanford and Livingston.
How does it come about that we are claiming an observation of the
spin 2 signature of the GW150914 gravitational wave that it
seems was missed by the LIGO/Virgo Collaboration team?
Because usually they apply a pre-processing filtering technique
that suppresses many possible low frequency contributions in the 
signals\cite{TheLIGOScientific:2016qqj,TheLIGOScientific:2016uux}.
Instead we use a pre-processing filtering method,
presented in \cite{Moreschi:2019vxw}, that avoids deformation
of the signal at low frequencies.
This has allowed us to reveal more astrophysical signal
in the LIGO strains of the GW150914 event,
at earlier times\cite{Moreschi:2019vxw}, with lower frequencies.

To study the presence of a similar signal in two detectors,
without using the theoretically constructed templates,
we have developed a new measure, that we call $\Lambda$, 
which is somehow related to the 
likelihood ratio\cite{Helstrom75} calculation for the detection
of a known signal in a single detector, and that we describe
below.

The organization of this article is as follows.
In section \ref{sec:newmeasure} we present a new measure
to compare the content of signals in the strain of two
detectors.
In section \ref{sec:detection} we present the analysis that
conduces us to the detection of similar signals in the two
strains for the GW150914 event.
Section \ref{sec:lambda_of_shift} is devoted to the study
of our measure $\Lambda$ as a function of the time shift
between the two strains, which conduces us to the detection of
the graviational wave polarization spin 2 signature in the GW150914 event.
And in section \ref{sec:final} we include some final comments
on this work.

\section{New measure for the comparison of signals in two detectors}\label{sec:newmeasure}

In the literature one can find many approaches for the study of data that
intends to determine whether a known signal is present in the data;
however works considering to determine
whether an unknown signal is present in two separate and independent
sets of data are rare.
In order to build the needed new measure, we have been 
guided by the usual method of maximum likelihood
as much as we could; 
since: ``Although the maximum likelihood principle is not based on any clearly defined
optimum considerations, it has been very successful in leading to satisfactory
procedures in many specific problems.''\cite{Lehmann2005}
The choice of a maximum likelihood method is also based on the fact
that we are dealing with a non-parametric detection\cite{Helstrom75}.
Due to the fact that from our proposed measure $\Lambda$, we will deduce important results,
we will go through a detailed presentation of it.
For this reason, in appendix \ref{sec:apendix-1}, we present a line of arguments
that support our choice for the new measure $\Lambda$.
These arguments are based on trying to adapt the likelihood method,
of searching for one signal in one data, to our case.
However we have to moderate the first natural definition of likelihood
so that the measure becomes useful.

The arguments presented in appendix \ref{sec:apendix-1}, are not intended to
be a deduction. They are only presented to connect our definition
of the measure $\Lambda$ with other related constructions that have been
used in related works. 
That is, there is no right or wrong definition of a measure,
any definition of a measure is arbitrary and so is ours;
the question is if it is useful for some purpose.
We claim that our definition is useful since 
it is  most powerful\cite{Lehmann2005} when compared
with two other choices and
it has allowed
us to detect the gravitational polarization in
the LIGO data of the GW150914 event, that we present below.

To give perspective to the strengths of our measure $\Lambda$, we here,
in section \ref{subsec:likelihood},
present our calculation of the likelihood ratio $\mathbf{L}$ to test the hypothesis
that a similar signal is recorded in the strains of two detectors,
versus the hypothesis that no signal has been recorded in both detectors;
we present our measure $\Lambda$ in section \ref{subsec:Lambda}
and we also recall the correlation coefficient $\rho$ between the two strains
in section \ref{subsec:correlation}.
Then, in section \ref{sec:detection} we discuss the properties of applying these
three measures to the case of the data of the GW150914 event.

\subsection{The likelihood ratio for the detection of an unknown signal in two detectors}\label{subsec:likelihood}

In appendix \ref{subsec:unknown-signal-two-detectors} we have deduced
the expression for the likelihood ratio to test the hypothesis
that a similar signal is recorded in the strains of two detectors,
versus the hypothesis that no signal has been recorded in both detectors,
which, in terms of the data, is given by:
\begin{equation}\label{eq:like-2t}
\begin{split}
\mathbf{L}( \mathbf{v}_1, \mathbf{v}_2) =& 
\exp \Bigg[\frac{m-1}{2} \Big(  \frac{ 1 }{ \sum_{k=1}^m v_{(1)k}^2 } 
+ \frac{ 1 }{ \sum_{k=1}^m v_{(2)k}^2 } \Big)
\sum_{k=1}^n v_{(1)k}\, v_{(2)k}
\Bigg]
;
\end{split}
\end{equation}
where the width of the window to calculate sample variances,
that is $m$,
is chosen appropriately depending on the nature of the observations,
and we are assuming that the means are zero.

This estimation of the desired measure have some difficulties.
The factor $\Big(\frac{ 1 }{ N_{0_1}} + \frac{ 1 }{ N_{0_2}} \Big)$(See
appendix \ref{subsec:unknown-signal-two-detectors} for details.)
is rather huge, when using LIGO data,
and it does not contribute to strengthen the
comparison of the data.
Expression \eqref{eq:like-2t} is the theoretical deduction of the
likelihood ratio, but for actual numerical application to the
gravitational wave data, we will use  \eqref{eq:like-3t}; for reasons
that we explain below.
Huge exponents are undesirable since might lead to unwanted
numerical error or even overflow errors.
For these reasons, we present our measure $\Lambda$ next.

\subsection{The measure $\Lambda$}\label{subsec:Lambda}

Due to the difficulties found in using the likelihood ratio,
just presented, we constructed a new measure that
turns out to be useful for our purposes.
Our arguments, that led us to this construction,
are presented in appendix \ref{sec:apendix-1}.

We here present the measure in synthetic form.

When dealing with data from gravitational wave observatories, one is confronted
with time series, which are supposed to contain signals from gravitational waves;
that might last, from a fraction of a second to several seconds;
depending on: the astrophysical nature of the source, the intensity,
noise state of the detectors at the time of recording, etc.
Then, in designing tools for the analysis of the data, 
and considering the transitory nature of the expected gravitational wave signals,
it is convenient
to introduce windowing techniques that allow for the study of portions
of the data.
For this reason we introduce an inner product that contemplates this point.

We define an inner product for two strains $\mathbf{x}(\tau)$ and $\mathbf{y}(\tau)$.
The inner product is calculated through the convolution with
an appropriate window $w(t-\tau)$, and is defined by:
\begin{equation}\label{key}
<\mathbf{x},\mathbf{y}>(t_j)
=
\sum_{k}  x(t_k)\, y(t_k) \, w(t_j - t_k) 
;
\end{equation}
where we use the fact that the data is obtained at discrete time
intervals.

For the specific case of the observation of gravitational waves at two detectors,
let $\mathbf{v}_1(\tau)$ represents the strain at one detector with respect to its proper time,
and $\mathbf{v}_2(\tau-\delta)$ represents the strain at the other detector, with time shift $\delta$.

Then, we define the time dependent $\Lambda$ measure from
\begin{equation}\label{eq:like-symmetric}
\begin{split}
\Lambda(\mathbf{v}_1,\mathbf{v}_2, \delta, t) =& 
\exp 
\Bigg[
\frac{1}{\sigma^*} 
\frac{ < \mathbf{v}_1 , \mathbf{v}_2 >
}{ 
	< ( \mathbf{v}_1 -\mathbf{v}_2), ( \mathbf{v}_1 -\mathbf{v}_2) > }   
\Bigg]
;
\end{split}
\end{equation}
where $\sigma^*$ is the standard deviation of 
$\frac{ < \mathbf{v}_1 , \mathbf{v}_2 >
}{< ( \mathbf{v}_1 -\mathbf{v}_2), ( \mathbf{v}_1 -\mathbf{v}_2) > }$.
In appendix \ref{sec:apendix-1} we present arguments that 
support this choice for the measure.

In short,
this measure can be understood as coming from an adaptation of
the likelihood method to our case,
with contrast accentuation and overall moderation.
Our choice gives reasonable results with actual LIGO data.

Although the arguments presented in appendix \ref{sec:apendix-1} use the idea
of having the same signal in two strains, the measure $\Lambda$ 
also indicates the existence of similar signals in two strains;
namely, when the signals $s_1$ and $s_2$ recorded in the detectors
satisfy $s_2 = s_1 + \epsilon$, with $\max|\epsilon| \ll \max|s_1|$.
This is precisely what happens when we apply the measure
to the data of the GW150914 event; as shown below.

\subsection{The correlation coefficient}\label{subsec:correlation}
One could have thought that the natural thing to do was just to consider
the correlation coefficient\cite{Ferguson67,McDonough1995,Casella2002,Lehmann2005} 
between the two strains,
with a time shift added to one of them.
Then, taking into consideration the characteristics of the gravitational waves
observation, mentioned above, we define the correlation coefficient in terms
of the natural inner product by:
\begin{equation}\label{eq:corr-coeff}
\rho_{\mathbf{v}_1 , \mathbf{v}_2} \equiv  \frac{ < \mathbf{v}_1 , \mathbf{v}_2 >
}{ \sqrt{< \mathbf{v}_1 , \mathbf{v}_1 > < \mathbf{v}_2 , \mathbf{v}_2 > } } 
.
\end{equation}
The fact is that the correlation coefficient gives very low response;
as we show in appendix \ref{sec:apendix-2-correlation} and in discussions below
in section \ref{sec:detection}.

\subsection{Minimal test for Gaussianity behavior}
In the preceding subsection we have presented a new measure to compare similar
signals in two detectors.
The motivating arguments are presented in the appendix \ref{sec:apendix-1}
which make use of the assumption that the noise
at both detectors are close to a Gaussian behavior.
Just by observing the amplitude spectral density of the strains at Hanford
and Livingston, one can see
that the data is no perfectly Gaussian.
However, there are many ways in which this assumption can be tested, and several
articles in the past have address this point;
but we here instead present graphs of what can be considered
the most direct minimal test for Gaussianity behavior,
namely the histogram of the data of the GW150914 event.
\begin{figure}[h]
\centering
\includegraphics[clip,width=0.45\textwidth]{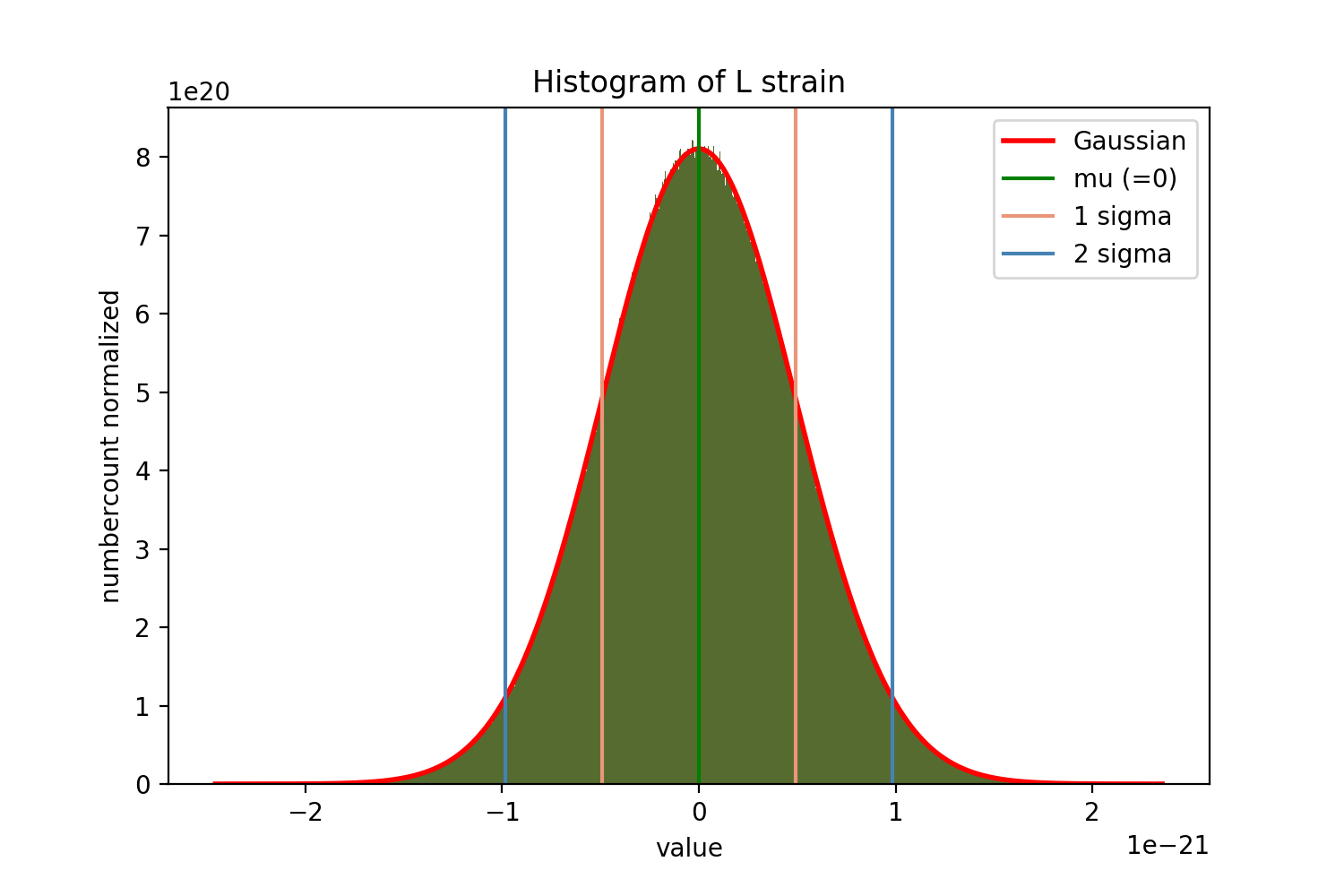}
\includegraphics[clip,width=0.45\textwidth]{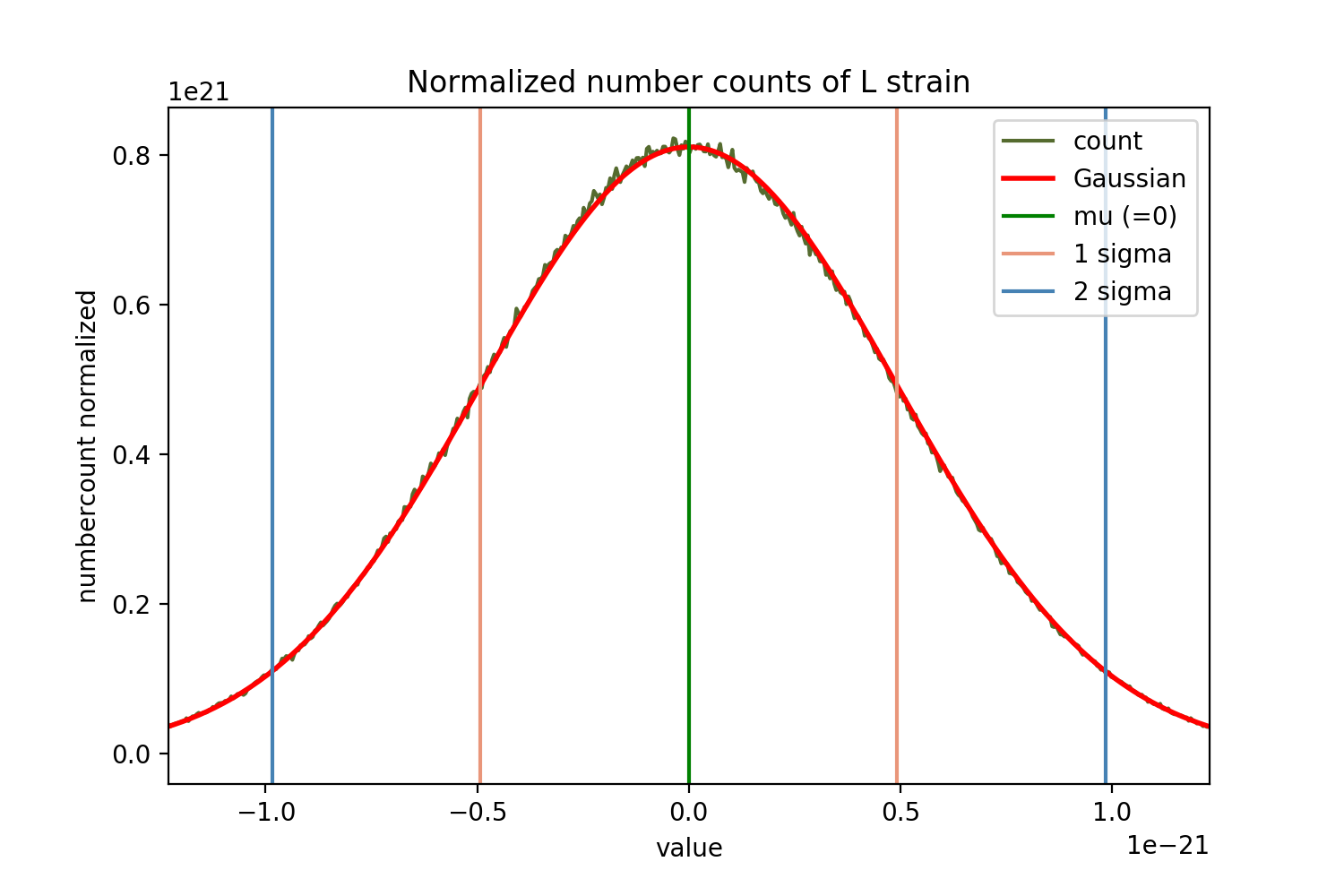}
\caption{Comparison of the histogram of Livingston data(green) with a Gaussian(red).
On the left the complete histogram of 256 seconds of the data, centered
at the time of the event. 
On the right the detail of the graph for $\pm2.5\sigma$. }
\label{fig:gaussianity_L-1}
\end{figure}
\begin{figure}[h]
\centering
\includegraphics[clip,width=0.45\textwidth]{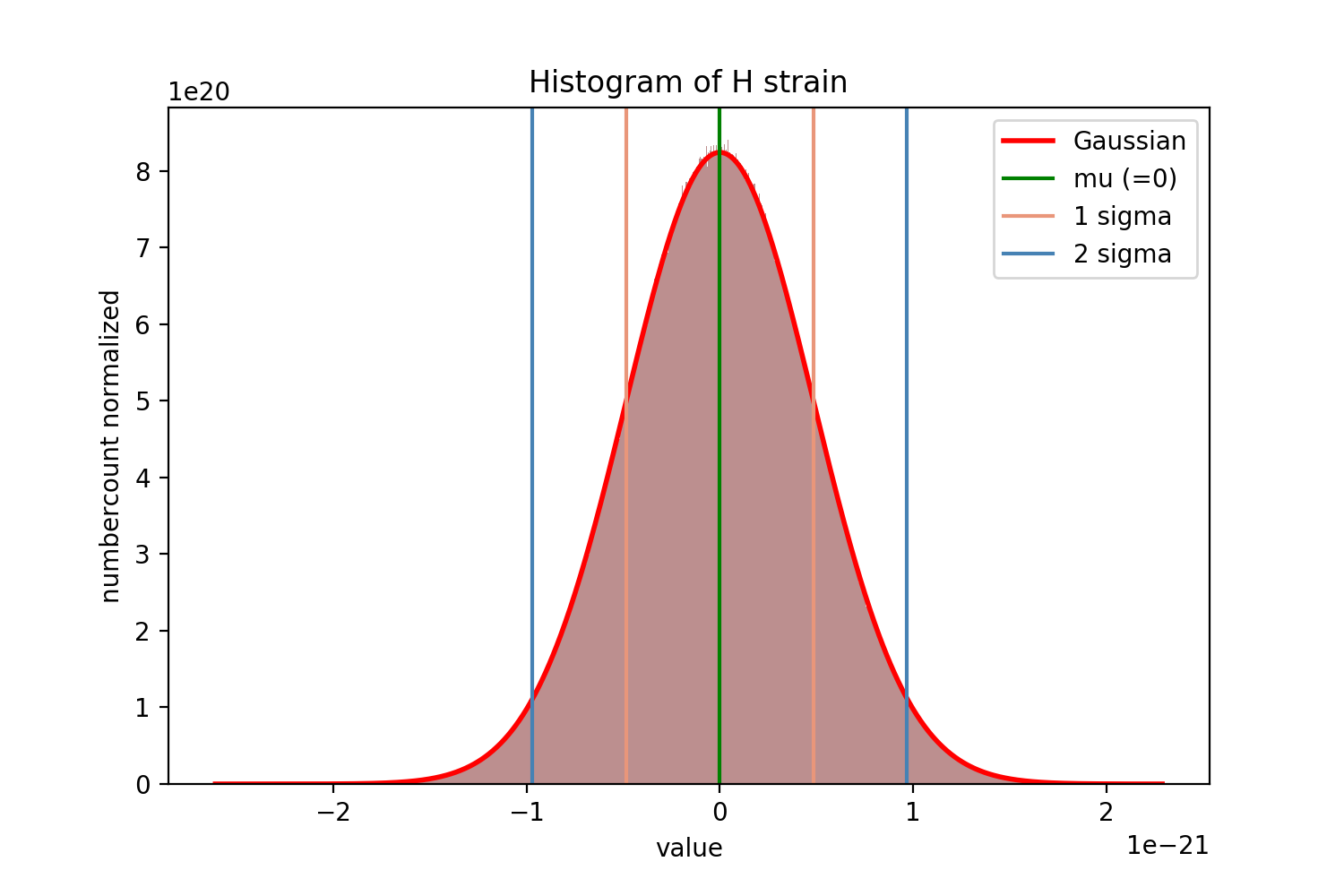}
\includegraphics[clip,width=0.45\textwidth]{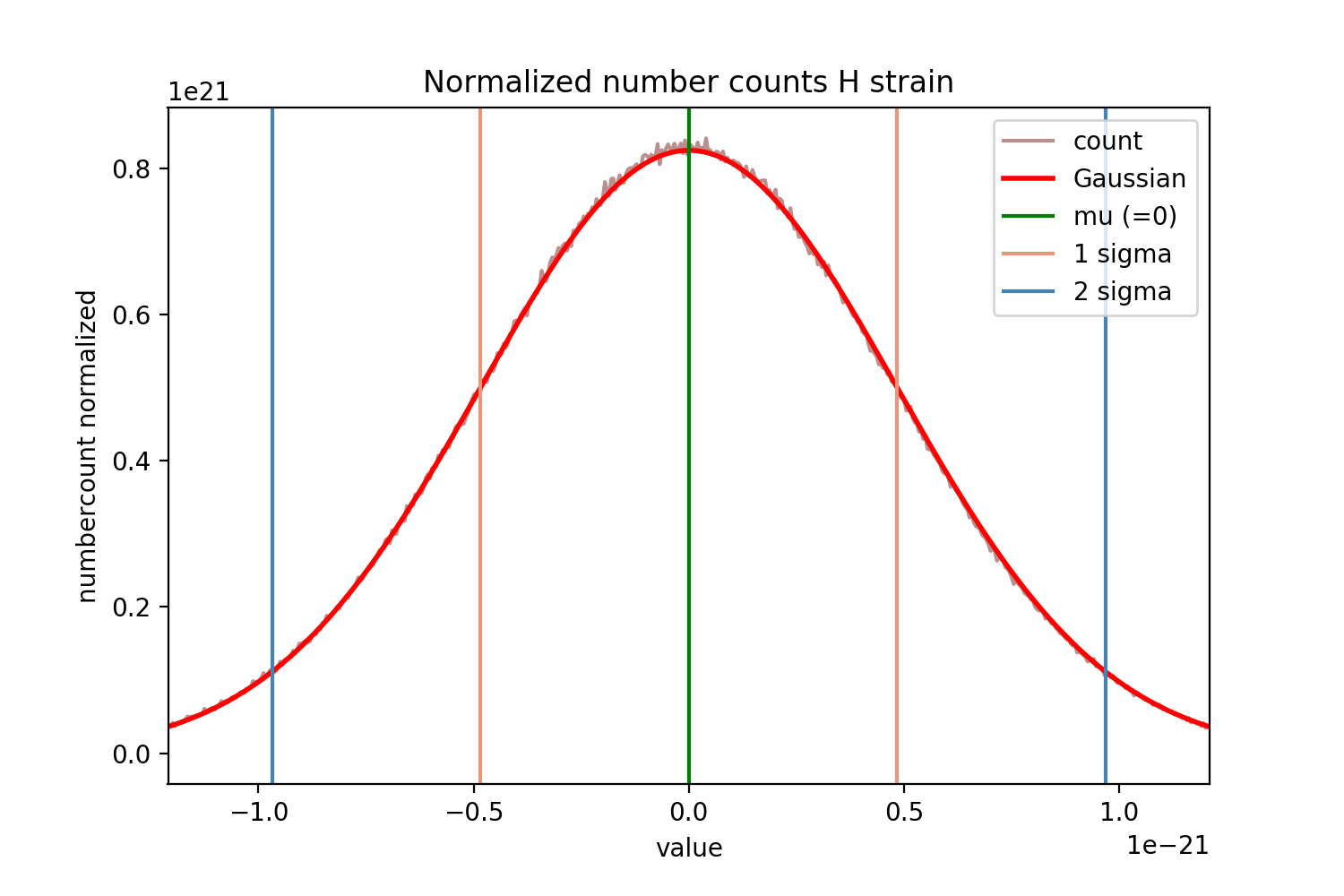}
\caption{Comparison of the histogram of Hanford data(colored surface) with 
	a Gaussian(red curve).
	On the left the complete histogram of 256 seconds of the data, centered
	at the time of the event. 
	On the right the detail of the graph for $\pm2.5\sigma$.}
\label{fig:gaussianity_H-1}
\end{figure}
We must emphasize that the behavior shown in the graphs of
figures \ref{fig:gaussianity_L-1} and \ref{fig:gaussianity_H-1},
are obtained after we have applied the pre-processing filtering
techniques that we have described in \cite{Moreschi:2019vxw}.
Figure \ref{fig:gaussianity_L-1} shows the histograms for the Livingston data
of 256 seconds centered at the time of the event,
and compared with a Gaussian with $\sigma_L=4.92212e-22$.
Figure \ref{fig:gaussianity_H-1} shows the histograms for the Hanford data
of 256 seconds centered at the time of the event,
and compared with a Gaussian with $\sigma_H=4.83976e-22$.
The standard deviations $\sigma_L$ and $\sigma_H$ were calculated
from the data.
For completeness, the Gaussian function is given by:
$G(x,\sigma,\mu)=\frac{1}{\sqrt{2 \pi} \sigma} \exp\big( -\frac{(x-\mu)^2}{2 \sigma^2} \big)$.
For both detectors the median was zero.

We would like to emphasize that although the Gaussianity properties of the GW150914 strain have been discussed
several times in the past, 
it is only when we apply our pre-processing filtering
techniques\cite{Moreschi:2019vxw} that this very clean Gaussian
behavior, shown in figures \ref{fig:gaussianity_L-1} and \ref{fig:gaussianity_H-1},
is achieved.

\section{Detection of similar signals in the two LIGO observatories}
\label{sec:detection}

\subsection{Study of the measure $\Lambda$ as a function of time for the GW150914 event}
\begin{figure}[H]
\centering
\includegraphics[clip,width=0.6\textwidth]{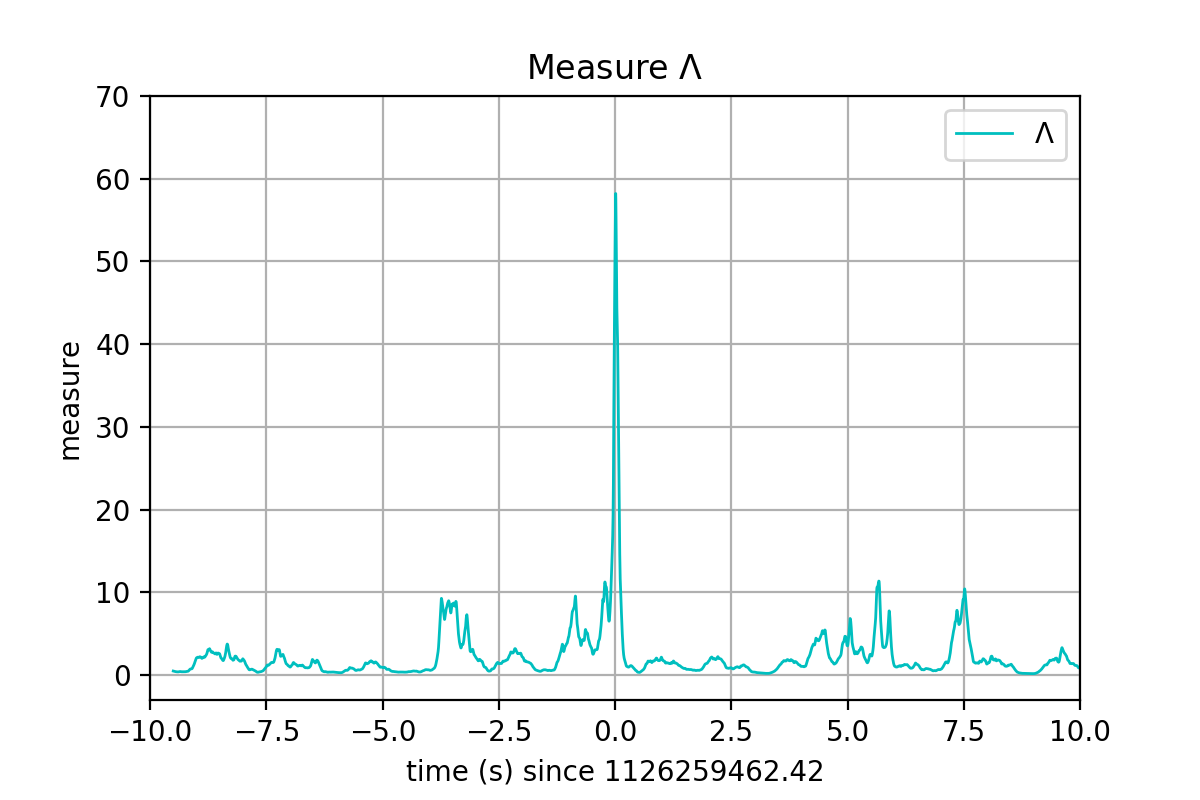}
\caption{The measure $\Lambda(t)$ for the shift $\delta=-0.007$s for
	the Hanford (-)strain, in the range $\pm10$s.
}
\label{fig:likelihood-1}
\end{figure}

We show in figure \ref{fig:likelihood-1} the graph of our 
measure $\Lambda$ as a function of time,
for the nominal Hanford(H) strain shift of $\delta=-0.007$s,  
in the interval -10s, 10s from the time of the event $t_e$;
for a Tukey window of 0.5s width,
where it can be seen that there is
a sharp peak at about 0.0s, that is at the time $t_e$;
where we have advanced the time axis by 
the width of the window.
The sign of the H strain is obviously chosen so that the natural
inner product with the L strain gives a positive result when correlating.
Figure \ref{fig:likelihood-1} shows that close to the time of the event there is a sharp
peak, indicating that most probably both detectors have recorded
a similar signal in the window $w$ before this time.

The sharp behavior of this measure invites us to calculate a coarse estimate of
the level of significance by using directly Chebychev's inequality\cite{Casella2002},
that we recall next:
\begin{teo}
	Let $X$ be a random variable and let $\lambda(x)$ be a nonnegative function.
Then, for any $r>0$,
\begin{equation}\label{key}
P(\lambda(X) \ge r) \le \frac{\mathbf{E}(\lambda(X))}{r}
.
\end{equation}
\end{teo}
Here $P$ means probability and $\mathbf{E}$ expectation value.

The first estimate of the level of significance $\alpha_0$ can be calculated from
identifying $\lambda=\Lambda$ and taking $r=\max(\Lambda)=58.21$.
Then, from the numerical calculation 
$\mathbf{E}(\Lambda)=2.338$,  we obtain $\alpha_0= 0.04016$.

With the knowledge of the standard deviation, we can use the customary form of
the Chebychev's inequality that sets:
\begin{equation}\label{key}
P(|X - \mu| \ge t \sigma_X) \le \frac{1}{t^2}
;
\end{equation}
where $\mu= \mathbf{E}(X)$ and $\sigma_X$ is the standard deviation.
Then, by identifying $X$ with $\Lambda$, and using the calculated value of
$\sigma_\Lambda = 4.4406$, we obtain the second estimate of the level of significance $\alpha_1$
given by $\alpha_1 = 0.0063$.
It can be seen that this second estimate improves on the first one;
since we are using more information on the statistics of $\Lambda$.

The Chebychev's inequality is normally applied when the probability distribution
for the situation under study is not known.
Let us try to infer more information on the statistical properties of $\Lambda$.
\begin{figure}[H]
\centering
\includegraphics[clip,width=0.6\textwidth]{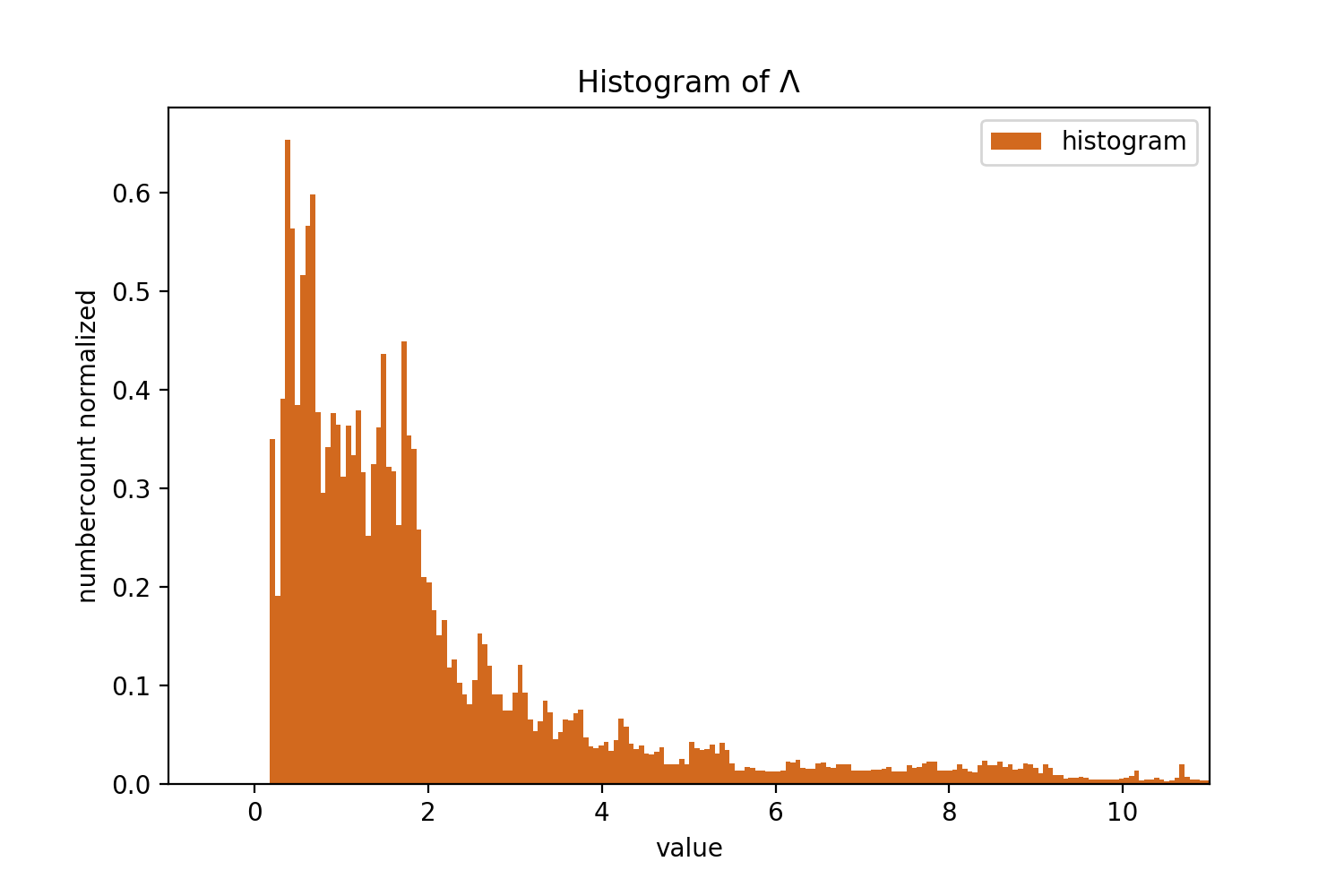}
\caption{Histogram of $\Lambda$ measure for a limited part of its domain.
}
\label{fig:histogram-Lambda-1}
\end{figure}
In figure \ref{fig:histogram-Lambda-1}, in order to show some detail, we present
the graph of the histogram of our measure for a limited part of its domain,
that does not include the maximum value, close to sixty.
The sample corresponds to a strain of 22 seconds centered at the time
of the event.
It can be seen, since the distribution only involves positive values,
that the histogram does not show a Gaussian behavior.
Instead it does resemble a log-normal behavior.

This, in turn invites us to see what is the behavior of the histogram
of the exponent, that we show next.
\begin{figure}[H]
\centering
\includegraphics[clip,width=0.6\textwidth]{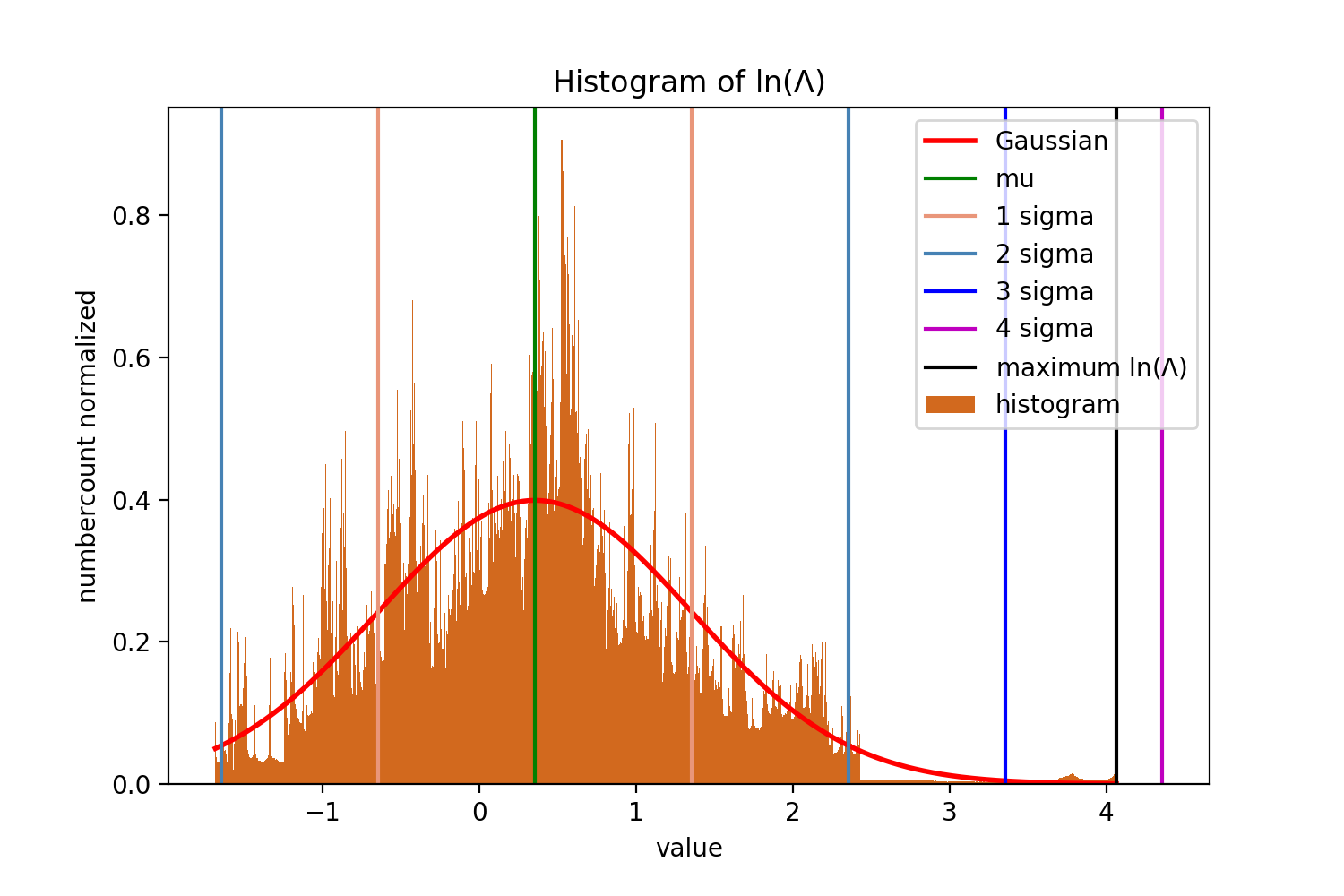}
\caption{Histogram of logarithm of $\Lambda$ measure.
It is also shown with verticals lines the median, 1, 2, 3 and 4 sigmas,
and with a black vertical line, the position of the maximum observed.
}
\label{fig:histogram-logLambda-1}
\end{figure}
In figure \ref{fig:histogram-logLambda-1} we show the graph of the histogram
of the logarithm of our measure for the 22s strain; where sigma is
the standard deviation.
The Gaussian curve is calculated from the mean and the standard deviation.
It can be observed that the Gaussian red curve is a good smoothed
approximation of the behavior of the histogram.
The black vertical line shows the position of the maximum of the measure,
observed at $z \sigma$, with $z = 3.70749$.

Identifying a Gaussian behavior for the logarithm of $\Lambda$,
we can assign the level of significance\cite{Lehmann2005}
$\alpha = (1/2) [\mathsf{erfc}(z/\sqrt(2))]$;
where $\mathsf{erfc}$ is the complementary error function.
Equivalently we can also define the confidence level $(1-\alpha)$.
The values so obtained are: a level of significance $\alpha = 0.00010466$, 
and a confidence level, or confidence coefficient\cite{McDonough1995}, of $(1-\alpha)= 0.99989534$. 

This is a remarkable strong behavior of the measure $\Lambda$;
since for the data we are analyzing it gives us 99.99\% confidence that
there are similar signals in both detectors, for the chosen window.
We will see next that the  measure $\Lambda$ is stronger than
the other two we are considering in this article.

\subsection{Study of the other measures as a function of time}

\begin{figure}[H]
\centering
\includegraphics[clip,width=0.45\textwidth]{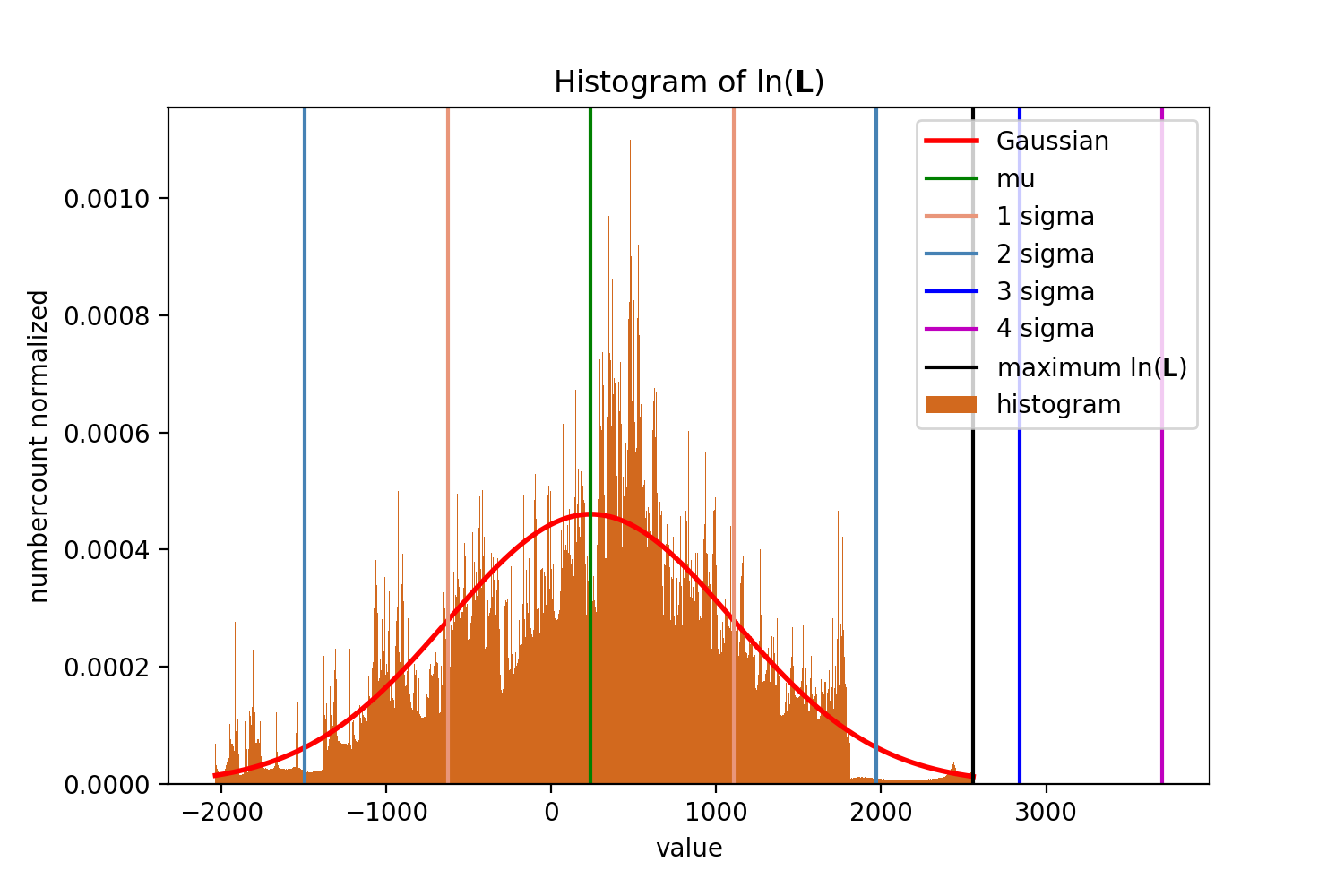}
\includegraphics[clip,width=0.45\textwidth]{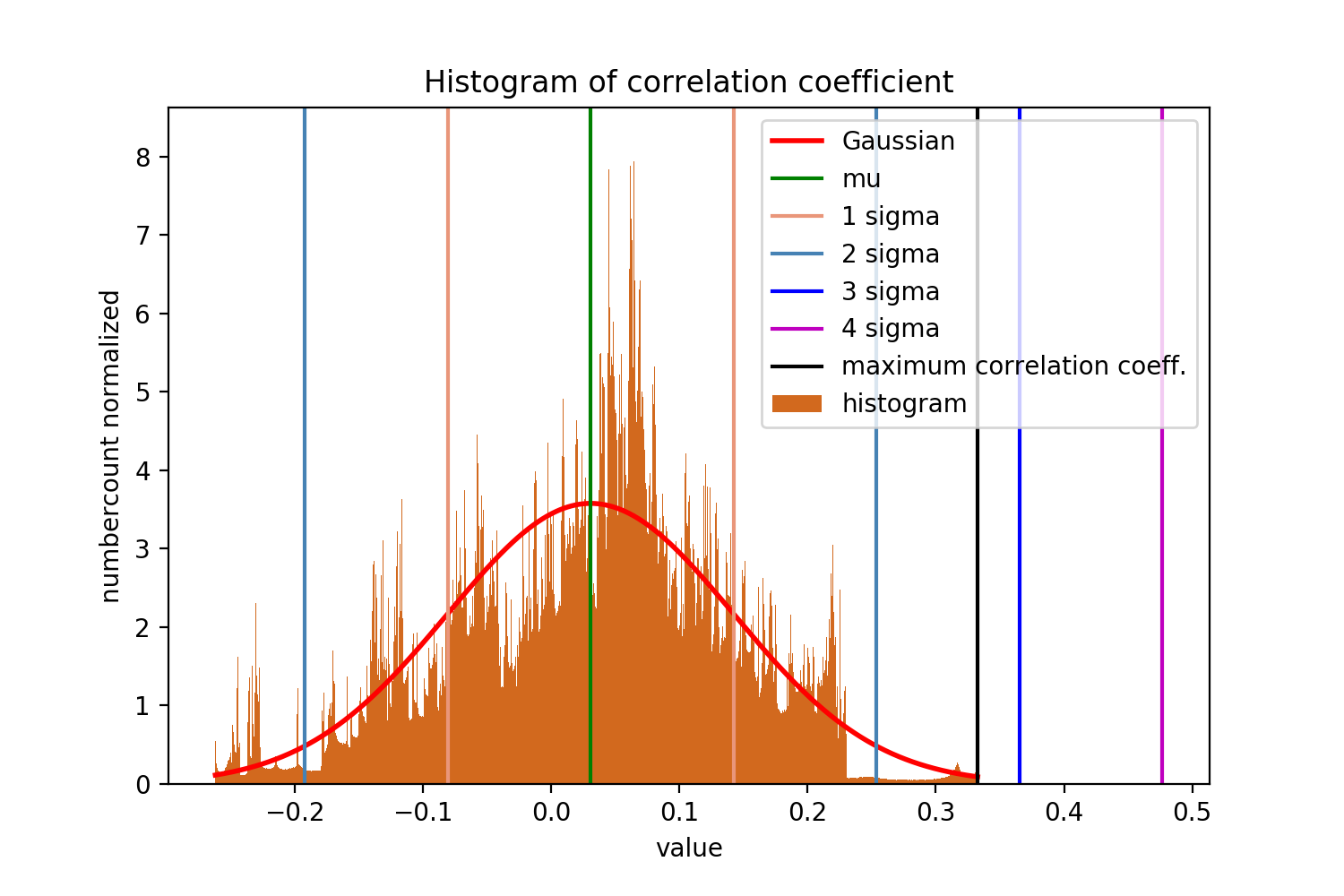}
\caption{Histogram of the logarithm of likelihood ratio $\mathbf{L}$
	and the correlation coefficient measure $\rho$.
	It is also shown with verticals lines the median, 1, 2, 3 and 4 sigmas,
	and with a black vertical line, the position of the maximum observed.
	They show a remarkable similarity in shape.
}
\label{fig:correl-coeffi-1}
\end{figure}

For the numerical calculation of the likelihood ration $\mathbf{L}$,
we make use of the natural inner product defined above,
so that the detailed expression of $\mathbf{L}$ to be used
with gravitational wave data is:
\begin{equation}\label{eq:like-3t}
\begin{split}
\mathbf{L}( \mathbf{v}_1, \mathbf{v}_2) =& 
\exp \Bigg[\frac{1}{2} \Big(  \frac{ 1 }{ < \mathbf{v}_1 , \mathbf{v}_1 >  } 
+ \frac{ 1 }{ < \mathbf{v}_2 , \mathbf{v}_2 > } \Big)
< \mathbf{v}_1 , \mathbf{v}_2 >
\Bigg]
.
\end{split}
\end{equation}
When we try to use the likelihood ration $\mathbf{L}$, that we calculated above,
to the data of the GW150914 event, we obtain an overflow error using python.
From the previous discussion, it is suggested to also study the behavior
of the logarithm of $\mathbf{L}$.
On the left hand side of figure \ref{fig:correl-coeffi-1} we shown the histogram
of the logarithm of the likelihood ratio, along with the position of
its median, sigmas, and the maximum of $\mathbf{L}$ with vertical lines
and the theoretical Gaussian calculated from the median and the standard deviation.
The maximum of $\mathbf{L}$ is located at $z_\mathbf{L} \sigma_\mathbf{L}$,
with $z_\mathbf{L} = 2.67756$ (Shown in figure \ref{fig:log-likelihood_007}.).
Identifying a Gaussian behavior for the logarithm of $\mathbf{L}$,
we can assign the level of significance 
$\alpha_\mathbf{L} = (1/2) [\mathsf{erfc}(z_\mathbf{L}/\sqrt(2))]= 0.00370803$.

Also, on the right hand side of figure \ref{fig:correl-coeffi-1} we shown the histogram
of the correlation coefficient, with the position of
its median, sigmas, and the maximum of $\rho$ with vertical lines
and the theoretical Gaussian calculated from the median and the standard deviation.
The maximum of $\rho$ is located at $z_\rho \sigma_\rho$,
with $z_\rho = 2.70806$.
Identifying a Gaussian behavior for $\rho$,
we can assign the level of significance 
$\alpha_\rho = (1/2) [\mathsf{erfc}(z_\rho/\sqrt(2))]= 0.00338$.

By comparing the level of significance that we can give to the detection of a similar
signal in the two LIGO observatories data, for the GW150914 event, 
using the three measures, we conclude
that the measure $\Lambda$, that we have introduced, is the strongest one.
For this reason from now on we just use $\Lambda$ as the working measure.

\section{Study of the measure $\Lambda$ as a function of shift for the GW150914 event}\label{sec:lambda_of_shift}

\subsection{The discovery of two local maxima}
\begin{figure}[H]
\centering
\includegraphics[clip,width=0.7\textwidth]{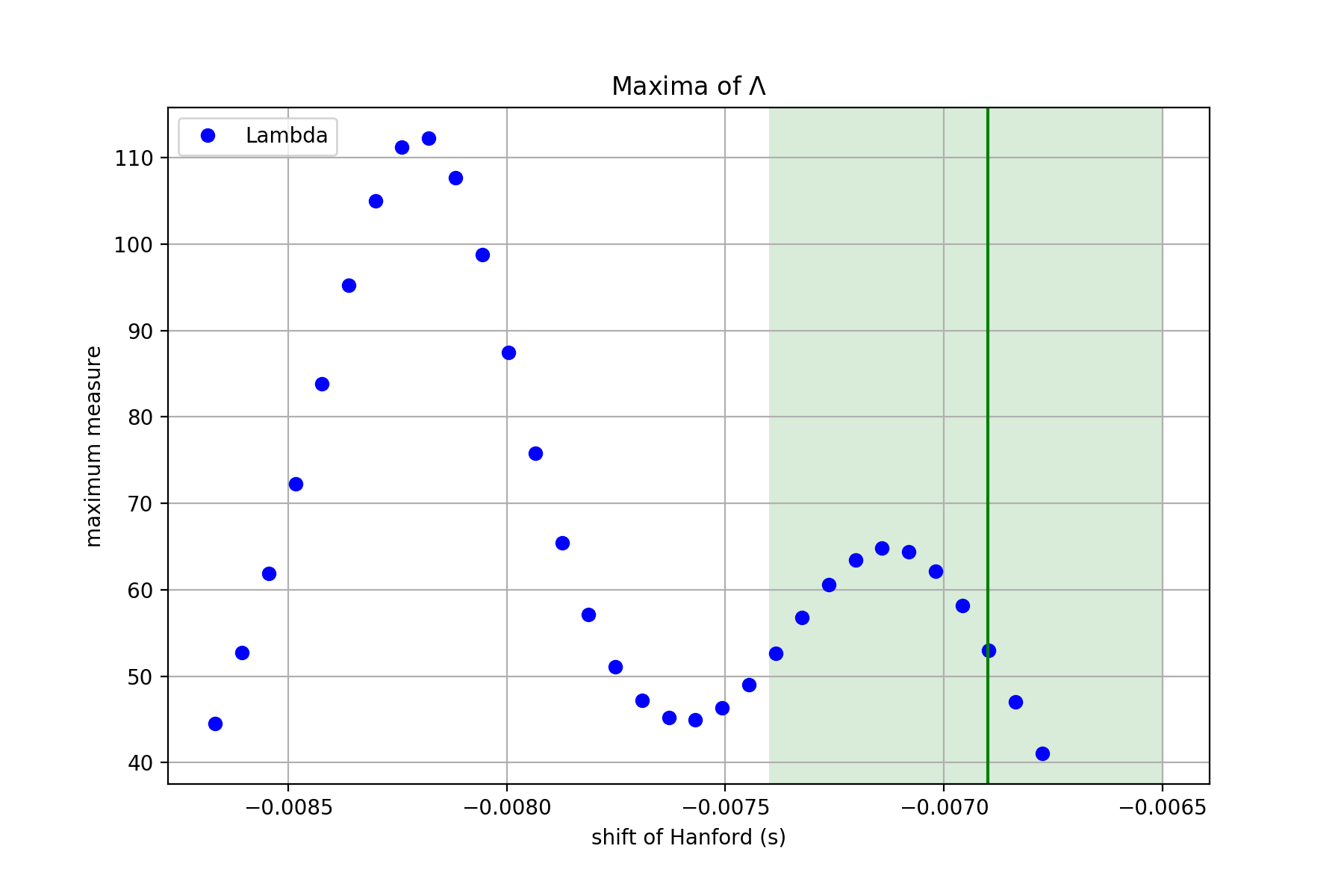}
\caption{The maxima of the measure as a function of the shift $\delta$ 
	for the (-)H strain, in sample steps,
	for a window of 0.5s width.
	The vertical line indicates the value of the shift suggested by 
	the LIGO/Virgo Collaboration for this event, and the vertical
	band indicates the error range as appear in their
	publications.
}
\label{fig:likelihood-2}
\end{figure}

In the graph of figure \ref{fig:likelihood-2} we show the detail
of the behavior of the maxima of the measure as a function of
a varying shift $\delta$ for the (-)H strain, in sample steps of 6.1035e-05seconds
in the range from -0.0087s to -0.0068s.
It is remarkable that in figure \ref{fig:likelihood-2}
one can see two local maxima instead of just one;
since one can see a peak at about -0.0071411s and another peak 
at about -0.0081787s, that is with an extra shift of -0.001037s.
We also indicate in figure \ref{fig:likelihood-2}, with a vertical line,
the value of the shift $6.9^{+0.5}_{-0.4}$ms suggested\cite{Abbott:2016blz,TheLIGOScientific:2016wfe} 
by the LIGO/Virgo Collaboration for 
the GW150914 event, along with its error range.
It can be seen that the peak on the right is in agreement
with the LIGO/Virgo estimate; but because of the
strong filtering techniques used by them, they could not
measure the second peak on the left of the graph.
The use of our tools for the appropriate determination of the
time shift between the two strains, will be carried out
elsewhere.

The appearance of two peaks in this kind of study is new
and we will interpret its physical meaning
in the next subsections.

\subsection{Fields, polarization of fields, polarization modes and spin weight quantities}
\label{subsec:fields-polarization}

In order to proceed with the discussion, it is convenient to distinguish first some concepts.

Let us say we are studying  the detection of a signal or excitation of some field.
To simplify the treatment we only consider that the field, carrying this excitation,
can be a scalar, a vector or a tensor field.

{\sf Polarization of the field:}
The polarization of the field is the specific characteristic nature of the
excitation, corresponding to the field.
For example, the polarization of an electromagnetic excitation, far away from
the sources, is a vector perpendicular to the direction of propagation.
The other example is the polarization of a gravitational excitation,
far away from the sources, which is normally described by a tensor 
whose main components are
taken with perpendicular vectors to the direction of the propagation.

{\sf Spin weight quantities:}
Having an excitation traveling along a null direction, it is convenient
to use a null tetrad basis, in which one of the vectors ($\ell^a$),
is chosen in the direction of the propagation; two complex 
(spacelike) null vectors ($m^a, \bar{m}^a$) are chosen perpendicular
to the propagation direction, and a fourth real null vector ($n^a$)
is chosen perpendicular to the last two and with
unit contraction with the first ($g_{ab} \ell^a n^b = 1$).
This is the basis for the GHP formalism\cite{Geroch73} which
is useful for the geometric discussion of a variety of situations
in which this type of basis appears naturally.
When contracting this basis with different objects one obtains
quantities with spin weight.
For instance, when having a vector $v^a$, the contraction with $m^a$
defines the quantity $\eta = g_{ab} v^a m^b$ which has spin weight 1;
instead $\bar\eta = g_{ab} v^a \bar m^b$ has spin weight -1.
If one is considering a tensor $t_{ab}$, one can obtain
components of spin weight -2, -1, 0, 1 and 2.

{\sf Polarization modes:}
The discussion of possible excitations in different  gravitational
theories that obey the \emph{principle of equivalence}\cite{Pirani64,Moreschi00}
led to the introduction of the notion of \emph{polarization modes}\cite{Eardley:1974nw}.
Analyzing plane gravitational waves excitations in \cite{Eardley:1974nw}
they found that there were six possible polarization modes:
real and imaginary parts of the Weyl component $\Psi_4$, which are
normally associated to the so called plus and cross polarizations,
having spin weight -2;
real and imaginary parts of the Weyl component $\Psi_3$, 
having spin weight -1;
and 
the Weyl component $\Psi_2$, having spin weight 0
and the Ricci component $\Phi_{22}$ which also has spin wight 0.
It has been customary in the literature to refer to to the  spin weight -1 quantity 
$\Psi_3$ as 'vector modes',
and to the  modes associated to the  spin weight 0 quantities
$\Phi_{22}$ and $\Psi_2$ as 'scalar modes', although
they should not be confused with
actual vector fields or scalar fields polarizations, discussed previously.

{\sf Behavior of the polarization modes in a general asymptotically flat spacetime:}
Suppose that we restrict further the consideration of gravitational theories to those
that, for isolated sources, admit asymptotically flat solutions.
Since, it is reasonably to think that the detectors of a gravitational wave perturbation,
located very far away from the sources, will observe the excitations on
the corresponding asymptotic background.
In \cite{Moreschi87} we have considered the general situation in which the
curvature of the spacetime decays to future null infinity with
a rate described by a function $f(\Omega)$, where $f$ is a positive monotonic
function such that $f(0)=0$, and $\Omega$ is the conformal factor needed in order
to define a continuous conformal metric at future null infinity.
This is the most general setting one can use to discuss asymptotically flat
spacetimes irrespective of any field equations.
Just from the algebraic condition of asymptotic flatness, one can deduce that
the components of the curvature go to zero at different rates; in particular
one can see that:
$\Psi_4 = f(\Omega) \hat{\Psi}_4 + \delta\Psi_4$,
$\Psi_3 = \Omega f(\Omega) \hat{\Psi}_3 + \delta\Psi_3$,
$\Psi_2 = \Omega^2 f(\Omega) \hat{\Psi}_2 + \delta\Psi_2$,
and
$\Phi_{22} = f(\Omega) \hat{\Phi}_{22} + \delta\Phi_{22}$,
where the hated quantities are regular at future null infinity
and the delta quantities go to zero faster than the previous
respective term.
If one also uses the differential structure encoded in the Bianchi identities,
in the interior of the spacetime,
then we can deduce more stringent relations, and one can see that actually
the asymptotic behavior must be of the form:
$\Psi_4 = - \Omega \ddot{\bar{\sigma_0}}  + O(-1)$,
$\Psi_3 = - \Omega^2 \eth \dot{\bar{\sigma_0}} + O(-2) $,
and
$\Phi_{22} = O(-1)$,
where $\sigma_0$ is a regular function at future null infinity,
a dot means time derivative, $\eth$ is a covariant angular derivative,
and a function $h(\Omega)$ is said to be order $O(q)$ if 
$\lim\limits_{\Omega\rightarrow 0} \Omega^q h =0$.
We deduce then, that from the possible excitations that could be contained
in the components $\Psi_4$, $\Psi_3$, $\Psi_2$ and $\Phi_{22}$, considered in \cite{Eardley:1974nw},
only the signals contained in $\Psi_4$ are measurable, if the sources
generate an asymptotically flat geometry.
Furthermore, one can even see that the leading order terms of 
$\Psi_4$ and $\Psi_3$ are not functionally independent.

To summarize, we could study the detection of gravitational waves in three scenarios.
\renewcommand{\theenumi}{(\roman{enumi})}%
\begin{enumerate}
\item We can consider very general kind of gravitational theories,
that might describe
gravitational excitations with scalar, vector and/or tensor fields.
\label{enu:uno}

\item We can consider gravitational theories,
that follow the principle of equivalence, but that might admit
solutions, for an isolated binary black hole system, which are not
asymptotically flat.
\label{enu:dos}

\item We can consider gravitational theories,
that follow the principle of equivalence, and whose  
solutions, for an isolated binary black hole system, are
asymptotically flat.
\label{enu:tres}
	
\end{enumerate}
In case \ref{enu:uno} we should determine whether the nature of the
gravitational excitation is due to a scalar field, vector field
or tensor field.

In case \ref{enu:dos} one is concerned to determine whether the 
modulus of the spin weight of the
gravitational tensor excitation is 0, 1 or 2.

In case \ref{enu:tres} we should check that the observations
of the gravitational excitations are consistent with
a tensor field nature and modulus of the spin weight 2 behavior.

In our view, case \ref{enu:tres} is the more plausible picture
for a system associated to the GW150914 event,
but we also use our findings in the situation
of case \ref{enu:uno} to infer some conclusions.
We do not consider the case \ref{enu:dos} in this work.

\subsection{Detection of the gravitational wave polarization}
In this subsection we would like to argue that the behavior found
in figure \ref{fig:likelihood-2} is precisely what one expects
when one records two polarization components of a gravitational wave.

The noise-free response $h(t)$ of one detector is given by
\begin{equation}\label{key}
h(t) = F_+ \, h_+(t) + F_\times \, h_\times(t) ,
\end{equation}
where $F_+$ and $F_\times$ are functions of the {\sf angles} describing
the orientation of the detector and the position and configuration of the source.

In order to have more insight into the detection process, it is useful to
have information of the expected functional form of the polarization components
of the gravitational wave, that one is hoping to observe.
Using basic physical models, the qualitative behavior of the expected signal
can be represented\cite{Sathyaprakash:1991mt,Cutler:1994ys} by:
\begin{equation}\label{eq:hdet}
h(t) = \frac{Q(\mathsf{angles})}{(t_f - t )^{1/4} }  
\cos\bigg(\phi_f(\mathsf{angles}) - 2 \big[\frac{t_f - t}{5 t_c}\big]^{5/8} \bigg)
.
\end{equation}
It can be seen that amplitude and frequencies are function of $(t_f-t)$,
and that the angular dependence will show as a modulation in the amplitude
and as a change of phase\cite{Sathyaprakash:1991mt}.
\begin{figure}[H]
\centering
\includegraphics[clip,width=0.45\textwidth]{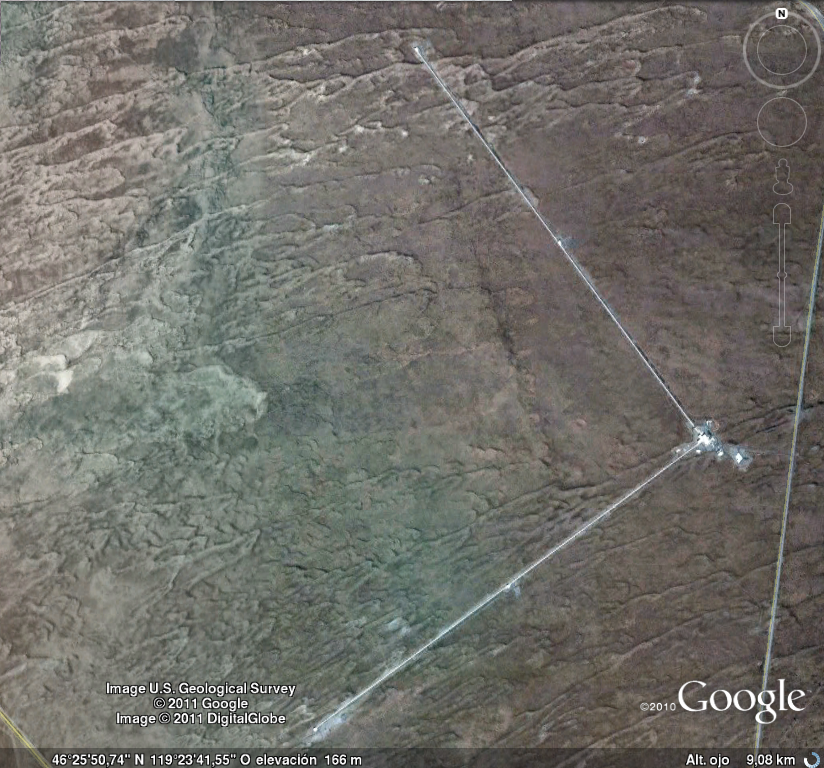}
\includegraphics[clip,width=0.45\textwidth]{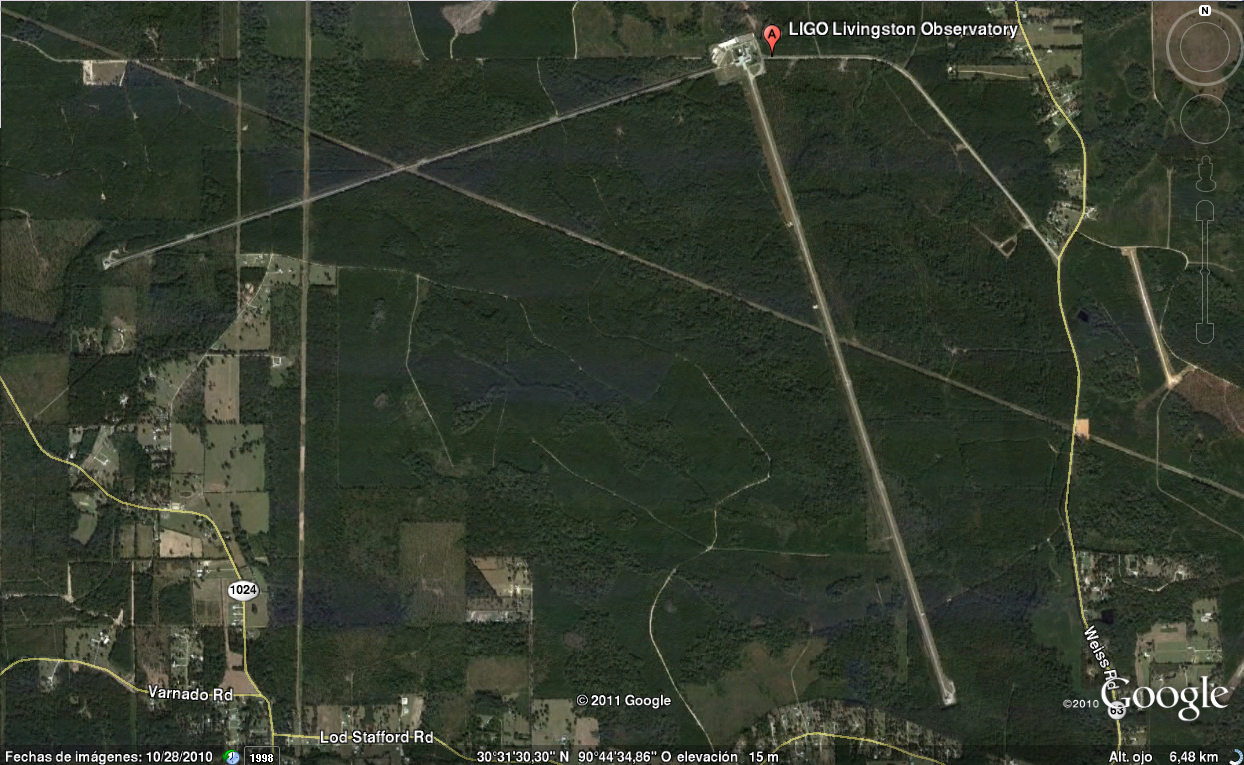}
\caption{
Satellite pictures of both detectors; where one can see a slight
different anti orientation
with respect to standard angular coordinates.
Of course the discrepancy increases when taking into account the curvature
of the Earth and the fact that they are separated by about 3000km.
}
\label{fig:observatories}
\end{figure}
In figure \ref{fig:observatories} we show the public satellite pictures
of both LIGO detectors; where one can see a slight different
anti-alignment of the arms, with respect to the terrestrial
angular coordinates. It is curious that in several 
LIGO publications\cite{Abbott:2018utx,Abbott:2017oio}
they assert that this configuration 
prevents Advanced LIGO from sensitively measuring more than a
single polarization mode; neglecting that even using
terrestrial angular coordinates they are not perfectly 
anti-aligned, and that due the the curvature of Earth
and the about 3000km separation, the orthogonal directions
are considerably different.

Dissimilar orientations of the detectors with respect to the signal
will not change the time position of the highest amplitude,
which occurs at the highest frequencies,
but will affect with a change of phase the low frequency
signals.
Therefore, if one is detecting the signal in a rather long
interval of time, the maximum of the measure $\Lambda$,
as a function of the shift $\delta$, will not coincide
with the value $\delta_e$ that marks the time shift
between the two detector times at the maximum of the signal,
i.e. the time of coalescence,
but it will be located at the $\delta$ needed to make
the coincidence of phases at lower frequencies, 
where several cycles contribute to the measure;
and therefore where more energy of the signal
might show agreement.
This is an effect that it has not been reported previously and is
a clear indication of the nature of the spin weight 2 character
(tensor polarization)
of the gravitational wave signal.
The fact that two peaks appear is accentuated by 
the lack of detection from the Livingston site,
in the small lapse of time that goes from about
-0.2s to -0.12s\cite{Moreschi:2019vxw}.

For the purpose of testing further our claim, we have calculated the matched templates for each detector
using the procedure described in the python scripts published by LIGO. 
The matched templates for each detector, 
that we could call $h_L(t)$ for Livingston and $h_H(t)$
for Hanford, obtained from the selected gravitational wave template,
suggested by the LIGO/Virgo Collaboration,
are the best estimates one has from LIGO to represent the gravitational wave signal
contained in the strains.

The two maxima effect
can also be observed directly in the matched templates
for each detector.
\begin{figure}[H]
\centering
\includegraphics[clip,width=0.7\textwidth]{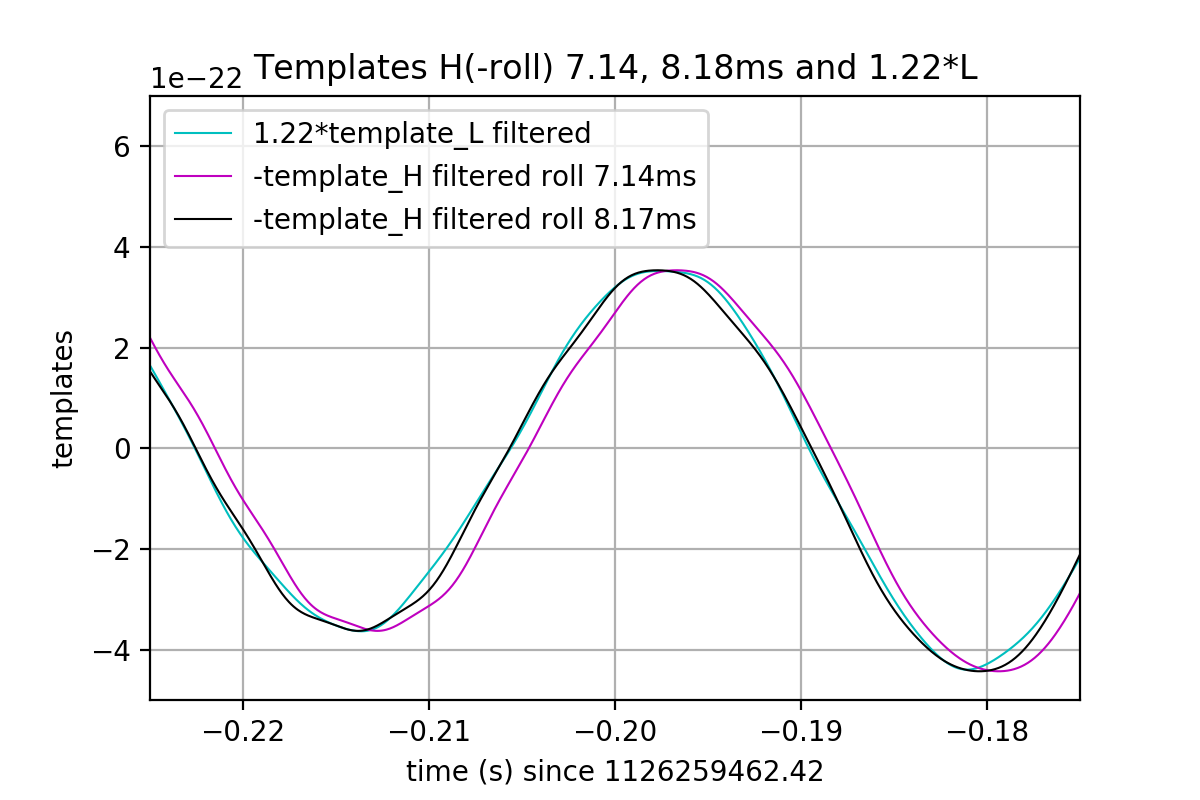}
\caption{The matched templates in the range -0.225s to -0.175s, 
	where the L matched template is shown in cyan and its has been rescaled by 1.22
	to achieve a better comparison.
	The (-)H template shifted to achieve the coincidence of the maximum with
	the corresponding L template maximum, occurring at about -0.005s, and so shifted 
	-7.14ms, 
	is shown in magenta.
	Finally 
	the (-)H template shifted to achieve the coincidence of the 
	second maximum of the Lambda measure, and so shifted -8.18ms, is shown in black.
	It can be seen that the shift necessary to go to the second maximum in Lambda
	is precisely the shift that obtains the coincidence of the templates
	of both detectors, in this range, at about -0.2s. 
	}
\label{fig:shifted-0,0012}
\end{figure}
This can further be noticed in the difference of the matched templates.
\begin{figure}[H]
\centering
\includegraphics[clip,width=0.48\textwidth]{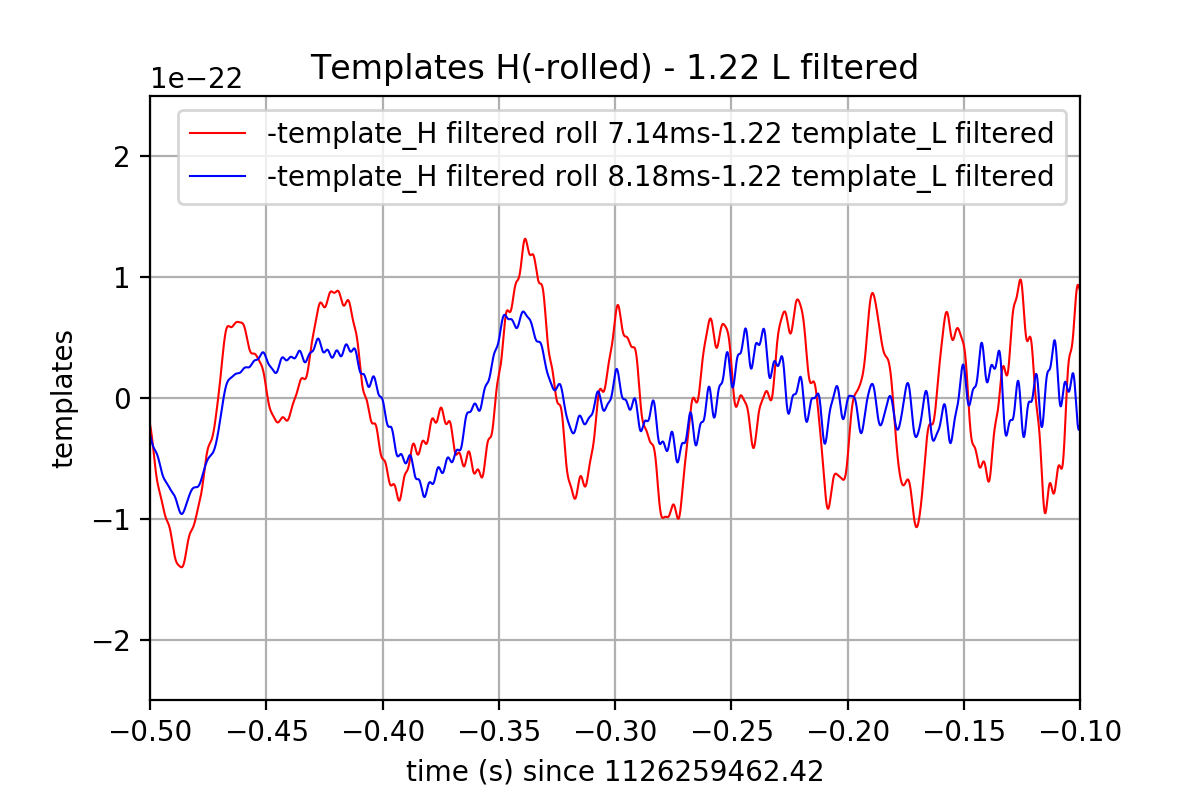}
\includegraphics[clip,width=0.48\textwidth]{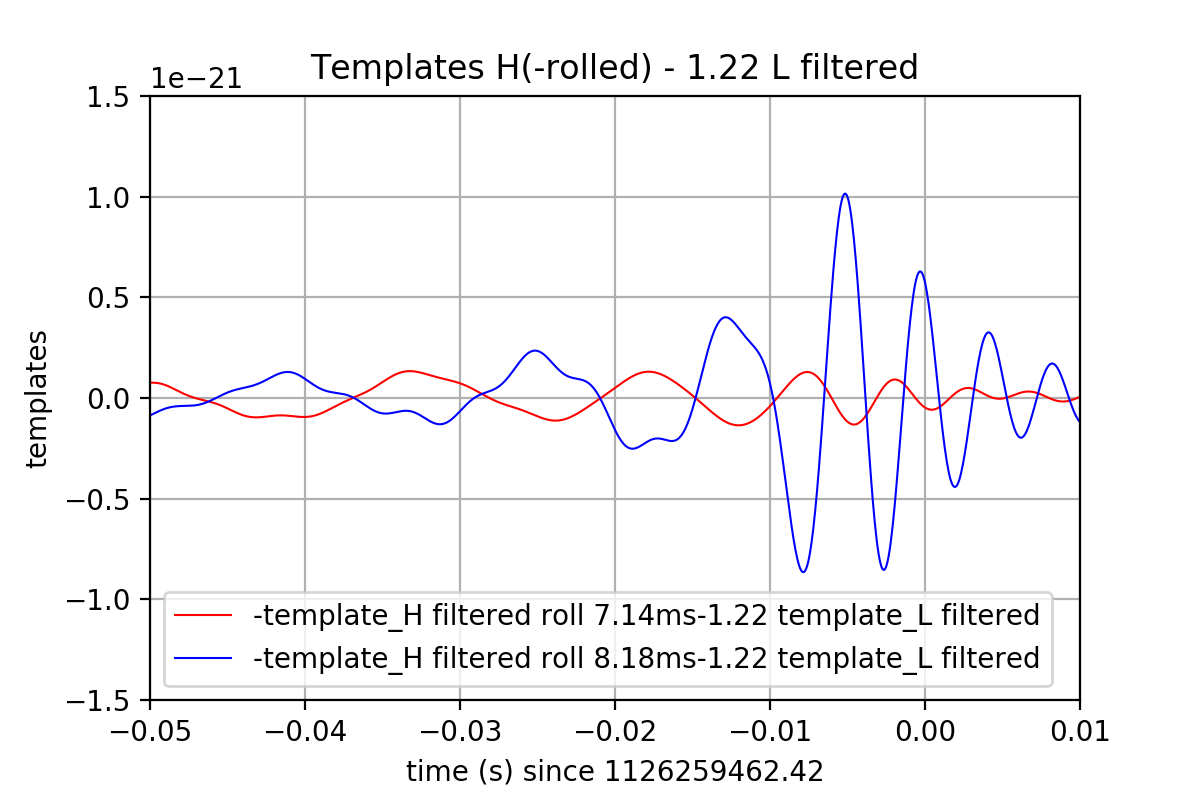}
\caption{On the left the differences of the matched templates for both detectors
	for the two shifts considered, in the range -0.5s to -0.1s.
	On the right also the differences, but in the range -0.05s to 0.01s.
	It can be seen that the shift of -7.14ms,
	shown in red, attains the minima
	differences in the range close to the time of the event.
	Instead, the -8.18ms shift, shown in blue, attains the minima differences
	in the range -0.5s to 0.1s.
	}
\label{fig:restas}
\end{figure}
Let us study the information encoded in the graph of figure \ref{fig:shifted-0,0012} in detail.
We show the matched template for the Livingston strain and 
the matched template for the Hanford strain with two different shifts, which are
obtained from the behavior of the $\Lambda$ measure as a function of the shift $\delta$
shown in figure \ref{fig:likelihood-2}.
It can be seen that 
an extra shift in the H strain, dictated by the second maximum in $\Lambda$,
attains an agreement (within the resolution of the graph) of phases at earlier times,
at around -0.2s, 
in both matched templates.
Instead when we apply the shift of -7.14ms, needed for phase agreement at 
the maximum amplitude, close to the time of coalescence, there is no agreement
in phase of the matched templates at times around -0.2s.
All this is reinforced by the graphs shown in figure \ref{fig:restas},
where on the left graph we show
the difference of the matched templates for both detectors
for the two shifts considered, in the range -0.5s to -0.1s;
and it can be seen that
the -8.18ms shift, shown in blue, attains the minima differences
in the this range.
Instead, on the right graph of figure \ref{fig:restas}
it can be seen that the difference of the templates for the
shift of -7.14ms,
shown in red, attains the minima
close to the time of the event,
in the range -0.05s to 0.01s.

Then, returning to the analysis of the observations by the two LIGO detectors,
we conclude that the first peak in $\Lambda$, shown in figure \ref{fig:likelihood-2},
 at about -0.00714s shift, 
is due to the concordance of both detected gravitational wave polarization
 signals of only the first 0.1s 
before the time of the event; where it has it highest local maxima
(in the time domain).
And that the second peak in $\Lambda$, at about -0.00818s,
is due to the extra shift needed in order to provide
agreement of phases of both gravitational polarization
signals at lower frequencies at earlier times.
Let us emphasize the this is only possible if each detector has
detected a different tensor polarization of the gravitational wave.

This is so because, if the signal where a vector excitation, then
each detector would at most observe a component of this vector,
that is the scalar product of the signal vector with the detector vector;
which is proportional to the norm of the vector excitation and
the angle with the detector vector.
Then, these two measurements will not introduce a 
relative different phase behavior, as the one detected in the GW150914 event.
And also, if the signal where a scalar excitation, then
each detector would observe the same signal, with a possible different
amplitude, due to state of the observatory. Again, these
processes will not introduce a different phase behavior, as
the one detected.

In other words, we deduce that the two LIGO strains for the GW150914 event,
contain the signal of two tensor polarization components
of the gravitational wave;
precluding any vector or scalar interpretation for the field 
transporting the excitation.

In this way we show that when the signal is strong enough,
one can detect the spin weight 2 nature of the signal
with just two detectors.

This finding is also useful for the determination
of the direction of the source in the sky, since
it gives more relations to be considered in
the angular calculations. We will study this in a
separate work.

It should be remarked that in reference \cite{Moreschi:2019vxw}
we have reported on the discovery of more
astrophysical signal in the GW150914 LIGO strains
than previously shown; based on the characteristics
of the signal in the time domain.
The finding we are reporting here,
is the second independent indication that there is
more physically interesting signal in the Hanford-Livingston strains
than those made known in previous LIGO/Virgo publications.

\section{Final comments}\label{sec:final}

Let us review the content of our presentation with its findings.
In this work we have presented a new measure that, we have
shown, is useful for the detection of similar signals in two
strains of gravitational wave observatories.
With this tool we were able to detect that the two LIGO strains
of the GW150914 event have a similar signal, close to the time
of the event, for about 0.5s;
and we have also detected that the two strains have
recorded two different components
of the polarization of the gravitational wave.
It is essential to stress first that we have used the pre-processing
filtering techniques, presented before\cite{Moreschi:2019vxw},
 as a starting point in the treatment of
the observed data at LIGO observatories.
Our filtering avoids the severe deformation that produces
the commonly used whitening techniques.
Our methods provide strains that have reasonable phase behavior;
solving some issues that have been noticed in the literature.
With this pre-processing filtering technique we have shown
in \cite{Moreschi:2019vxw} that there seems to be an astrophysical signal that
extends to about 0.5s previous to the time of the  GW150914 event.
This is much more than the 0.1s and 0.2s reported in LIGO articles.

In more detail, in section \ref{sec:newmeasure} we 
have presented a new measure $\Lambda$ to compare the existence
of unknown similar physical signals in two detectors.
And we have also presented the calculation of the likelihood ratio $\mathbf{L}$
for the detection of two similar signals in two strains.
Due to the fact that several authors use correlation 
coefficients\cite{Green:2017voq,Liu:2016kib,Creswell:2018tsr,Liu:2018dgm,
	Liu:2018whi,Nielsen:2018bhc,Jackson:2019xbq}
to analyze the observed data, we have also considered 
in section \ref{sec:newmeasure}
the corresponding correlation coefficient $\rho$ for
the two strains.
In appendix \ref{sec:apendix-2-correlation}
we have presented the result of applying the correlation coefficient
to the problem we have treated here; and we have shown
that it gives information with too much noise.
We have also shown that it is a weaker measure, when compared
with $\Lambda$ measure.
In fact, we have shown, that from the statistical consideration
of the three measures considered in section \ref{sec:newmeasure},
our measure $\Lambda$ can be used with a level of significance,
for this problem, which is more than 30 times stronger
than those than can be used with the likelihood ratio $\mathbf{L}$ or 
the correlation coefficient $\rho$.
For all those reasons,
we prefer the measure $\Lambda$,
for the purpose of this work.

The application of the measure $\Lambda$,
in section \ref{sec:detection}, to the two LIGO strains
of the GW150914 event, with a nominal time shift of -0.007s,
for (-)Hanford data, and with a window of 0.5s length,
gives a clean, sharp peak as a function of time;
which can be assigned a crude level of significance 
$\alpha_1 = 0.0063$, or a 
level of significance $\alpha = 0.00010$,
if we identifying a Gaussian behavior for the logarithm of $\Lambda$.
Another way of saying this is that we can trust the hypothesis
that there is a similar signal in both strains with a 
99.99\% confidence level.

When studying, in section \ref{sec:lambda_of_shift}, the behavior
of the measure $\Lambda$ as a function of the time shift between
the strains, we found two local maxima.
We have shown, that the existence of this effect, and
the time lapse between the local maxima, is completely
consistent with the comparison of two different tensor polarization
components of the gravitational wave one is expected
to receive for the GW150914 event.
In particular we have shown that the signals recorded in both
detectors are not the same.
Our measure allows to infer the spin 2 nature (tensor polarization)
of the gravitational waves detected in
the LIGO GW150914 event, with just two detectors.
The characteristics of the signals detected by both LIGO
observatories for the GW150914 event,
precludes scalar field and vector field interpretations,
in favor of the spin 2 tensor field description.

Note that in this work we have not attempted to determine the exact
functional form of the gravitational wave; since all the information
comes from the comparison of the signals through the use
of our measure $\Lambda$.

To our knowledge this is the first time that the gravitational
polarization has been determined by observation.
In fact, the LIGO/Virgo Collaboration has stated that
there were need of five 
detectors\cite{Abbott:2017tlp,Abbott:2018utx}
in order to be able
to determine the spin 2 nature of gravitational waves.
In particular, in \cite{Abbott:2018utx} they assert:
"...the LIGO-Hanford and LIGO-Livingston
detectors are nearly co-oriented, preventing Advanced
LIGO from sensitively measuring more than a single
polarization mode..."
If this were true, then the study we have carried out on the two
strains of the GW150914 event, would have shown only one
peak as a function of the time shift.
Instead, we have shown here that with our careful pre-processing
filtering techniques, and applying the $\Lambda$ measure,
we have found two peaks, and, as explained above,
we have detected that
the strains in the two LIGO observatories of the GW150914 event
have recorded two different components
of the tensor field gravitational wave.

Other researchers have tackled the problem of investigating the
existence of a signal in two detectors without recurring to
the use of a template.
For example in \cite{Liu:2018dgm} they have studied a
template-free method for determining the best common
signal in the Hanford and Livingston detectors for the GW150914 event.
They have assumed the existence of an exact same signal in
both detectors.
Their objective was to find such signal;
instead we concentrate only in trying to determine whether
there exist a similar signal in both detectors,
although we believe that our measure could be useful
to the objectives of \cite{Liu:2018dgm} and \cite{Liu:2018whi}.
Also it can be noticed that they have used different pre-processing
filtering techniques than we have done.

The implication of our detection of two local maxima
on the localization and configuration of the source of the GW150914 event,
involves a different kind of concepts than those analyzed here,
and will be studied elsewhere along with other implications
of our finding.

We expect that our measure will become useful in checking
possible gravitational lensed black hole mergers\cite{Haris:2018vmn}
by providing a new approach to this problem.


\subsection*{Acknowledgments}

My great thanks to Carlos Briozzo and Dante Paz for a careful reading
of the manuscript and for comments and discussions on statistics and
data analysis, and to Emanuel Gallo for criticism and discussions
on the physics of gravitational wave signals.

We are very grateful to the LIGO/Virgo Collaboration for making available the
data and the python scripts on data analysis
at \href{https://www.gw-openscience.org/}{https://www.gw-openscience.org/}.

We acknowledge support from  SeCyT-UNC, CONICET and Foncyt.


\appendix

\section{Arguments to build the measure}\label{sec:apendix-1}

\subsection{Detection of a known signal}\label{subsec:known-signal}
We here recall the basics of 
the likelihood method as applied to one set of data
following the notation of \cite{Helstrom75}.

Samples of white Gaussian noise
taken by an instrument having a bandwidth $W = 2\pi \Delta \nu$ 
will be Gaussian
random variables with the probability density function (p.d.f.)
\begin{equation}
p_1(x) = \frac{1}{\sqrt{2\pi N_0}} e^{-\frac{x^2}{2 N_0}}
,
\end{equation}
with
\begin{equation}
N_0 = \frac{N W}{2 \pi} = N  \Delta \nu 
;
\end{equation}
where $N$ is the unilateral spectral density.
The joint p.d.f. of samples $x_1, x_2, ..., x_n$ at times $t_1, t_2, ..., t_n$
separated by intervals much longer than $2\pi/ W$ 
will be statistically independent, given by
\begin{equation}\label{eq:gaussian-noise}
p(x_1,t_1;x_2,t_2;...;x_n,t_n) = 
\prod_{k=1}^{n} p_1(x_k) =
\frac{1}{(2 \pi N_0)^{n/2}} e^{\big( -\frac{1}{2 N_0} \sum_{k=1}^{n} x_k^2\big)}
.
\end{equation}
In treating the detection of signals in the presence of this kind of noise,
we shall imagine sampling the random processes by an instrument whose
bandwidth is much greater than that of any of the signals involved. We
can then apply eq. (\ref{eq:gaussian-noise}) to the values of the noise at times
$t_1, t_2, ..., t_n$ that are arbitrarily close.

Let $s(t)$ be a signal superimposed on a Gaussian noise $\mathsf{n}(t)$;
so that one observes
\begin{equation}
v(t) = \mathsf{n}(t) + s(t) ,
\end{equation}
in the interval $0< t < T$.
The hypotheses $H_0$ is that the signal is not present,
and the hypotheses $H_1$ is that the signal is present
in the observation $v$.

The observations are made at $n$ uniformly spaced times
$t_k = k \Delta t  = k \frac{T}{n}$, with $k=1,2,...,n$;
with values $v_k = v(t_k)$.
The observations for the two possibilities are described
by the joint probability density functions
$p_0(\mathbf{v}) = p_0(v_1,v_2,...,v_n)$ and
$p_1(\mathbf{v}) = p_1(v_1,v_2,...,v_n)$ under the hypotheses
$H_0$ and $H_1$ respectively.
The observer's decision is best made on the basis of the 
likelihood ratio, 
\begin{equation}\label{eq:likelihood_1}
\mathbf{L}(\mathbf{v}) 
= \mathbf{L}(v_1,v_2,...,v_n) 
= \frac{p_1(\mathbf{v})}{p_0(\mathbf{v})}
.
\end{equation}
Its value for the data at hand is compared with a fixed decision level 
$\mathbf{L}_0$; if $\mathbf{L}(\mathbf{v}) <\mathbf{L}_0$
the observer decides that there is no signal present.

It is assumed that
the measurements of $v(t)$ at times $t_k$ are made by
an instrument of such a large bandwidth that however small the intervals
$\Delta t$ between them, their outcomes have statistically independent 
noise components. Then, under hypothesis $H_0$ their joint 
probability density function (p.d.f.) is
\begin{equation}\label{eq:pdf_0}
p_0(\mathbf{v}) = 
\frac{1}{\big( 2 \pi N_0  \big)^{-n/2}}
\exp \Big( - \sum_{k=1}^n \frac{v_k^2}{2 N_0}  \Big)
,
\end{equation}
with $N_0 = \frac{N \, W}{2\pi}$.

When the signal is present, the part of the observed $v_k$ due to the noise is
$v_k - s_k$, with $s_k = s(t_k)$ a sample of the signal. 
Therefore, the data $v_k$ should behave as 
independent
Gaussian random variables with mean values $s_k$ and variances $N_0$,
namely,
under hypothesis $H_1$ the joint p.d.f. of the data is,
\begin{equation}\label{eq:pdf_1}
p_1(\mathbf{v}) = 
\frac{1}{\big( 2 \pi N_0  \big)^{-n/2}}
\exp \Bigg( - \sum_{k=1}^n \frac{(v_k - s_k)^2}{2 N_0}  \Bigg)
.
\end{equation}

The likelihood ratio, eq. (\ref{eq:likelihood_1}), now becomes
\begin{equation}
\mathbf{L}(\mathbf{v}) 
=
\exp \Bigg( \sum_{k=1}^n \frac{2 s_k\, v_k - s_k^2}{2 N_0}   \Bigg)
.
\end{equation}
The observer chooses hypothesis $H_0$ if 
$\mathbf{L}(\mathbf{v})  < \mathbf{L}_0$  or, because of the
monotone character of the exponential function, if
\begin{equation}
\Delta t \sum_{k=1}^n  s_k\, v_k 
<
\frac{1}{2} \Delta t \sum_{k=1}^n  s_k^2 
+ N_0 \Delta t \ln \mathbf{L}_0
.
\end{equation}
Hence he can base his decision on the value of the quantity
\begin{equation}\label{eq:G_n}
G_n =  \Delta t \sum_{k=1}^n  s(t_k)\, v(t_k) , 
\end{equation}
comparing it with some fixed amount $G_{n0}$ determined by some criterion. 
In the n-dimensional Cartesian space with coordinates $v_k$, 
the decision surface $D$ is a hyperplane
\begin{equation}
\sum_{k=1}^n  s(t_k)\, v(t_k)
= \text{constant}
,
\end{equation}
which is perpendicular to the vector with components $s_k$.

It can be seen that the natural inner product of the expected signal
and the strain, that we denote by 
$<\mathbf{v},\mathbf{s}> = \sum_{k=1}^n  s(t_k)\, v(t_k)$, 
is the basic quantity of the likelihood calculation.
And if $<\mathbf{s},\mathbf{s}>$ can be neglected in front
of $<\mathbf{v},\mathbf{s}>$, or if one only concentrates
in the functional dependence on the data, one arrives at the 
working expression for the likelihood to be
\begin{equation}
\mathbf{L}(\mathbf{v}) 
=
\exp \Bigg(  \frac{<\mathbf{v},\mathbf{s}>}{N_0}   \Bigg)
;
\end{equation}
where for the sake of simplicity in this presentation,
we are assuming a constant $N_0$, although the expressions
can easily be generalized.
Normally, $N_0$ is measured from the local properties of the
data, close to the time of the event under study.

In actual situations, in which the expected signal has some
characteristic length in time, one does not use the natural inner product
but a convolution of it with an appropriately chosen window $w$,
with a width of the order of the characteristic length of
the signal. So that one actually works with the definition:
\begin{equation}\label{eq:inner-app-1}
<\mathbf{v},\mathbf{s}>(t_j)
=
\sum_{k}  v(t_k)\, s(t_k) \, w(t_j - t_k) 
.
\end{equation}
In this way one samples the data with an appropriate width.

The likelihood method is used by the LIGO/Virgo Collaboration
as a standard way to obtain the matched templates to the
observed signals\cite{LIGOScientific:2019hgc,TheLIGOScientific:2016wfe}.

\subsection{The case of the same (similar) unknown signal in two detectors}\label{subsec:unknown-signal-two-detectors}
To simplify the notation we are going to omit when possible the index
denoting the time variation.
Let one detector observe the data $v_1$, which is supposed to contain the signal $s_1$
in the presence of the noise $\mathsf{n}_1$, and similarly for the other detector so that
\begin{equation}
v_1 = \mathsf{n}_1 + s_1 ,
\end{equation}
and
\begin{equation}
v_2 = \mathsf{n}_2 + s_2 ;
\end{equation}
but the basic assumption is that both detectors contain the same signal 
(Although it also applies to similar signals $s_2 = s_1 + \epsilon$, for some
small $\epsilon$, as we will show below.)
\begin{equation}
s_1 = s_2 = s
.
\end{equation}
Then one can express
\begin{equation}
\mathsf{n}_1 = v_1 - s = v_1 - v_2 + \mathsf{n}_2
;
\end{equation}
so that instead of (\ref{eq:pdf_1}) now we will have
\begin{equation}\label{eq:pdf_2}
p_1(\mathbf{v}_1) = 
\frac{1}{\big( 2 \pi N_{0_1}  \big)^{-n/2}}
\exp \Bigg( - \sum_{k=1}^n \frac{(v_{(1)k} - v_{(2)k} + \mathsf{n}_{(2)k})^2}{2 N_{0_1}}  \Bigg)
;
\end{equation}
where $\mathbf{v}_1=(v_{(1)1}, v_{(1)2},..., v_{(1)k}, ...)$ denotes 
the complete strain of detector 1.
In this situation, hypothesis 1 is that detector 1 have recorded the same signal as detector 2,
and hypothesis 0 is that detector 1 has not recorded the signal.

Similarly for detector 2 one also has
\begin{equation}\label{eq:pdf_3}
p_2(\mathbf{v}_2,s_2) = 
\frac{1}{\big( 2 \pi N_{0_2}  \big)^{-n/2}}
\exp \Bigg( - \sum_{k=1}^n \frac{(v_{(2)k} - s_{(2)k} )^2}{2 N_{0_2}}  \Bigg)
.
\end{equation}

Assuming the statistical independence
of the measuring process in the two detectors 
we arrive at the the joint probability from the product of the probabilities
calculated for each detector; namely
\begin{equation}\label{eq:joint-1}
\begin{split}
p(\mathbf{v}_1,s_1,\mathbf{v}_2,s_1)
=&
\frac{1}{\big( 2 \pi N_{0_1}  \big)^{-n/2}}
\exp \Bigg( - \sum_{k=1}^n \frac{(v_{(1)k} - v_{(2)k} + \mathsf{n}_{(2)k})^2}{2 N_{0_1}}  \Bigg)
\\
&
\frac{1}{\big( 2 \pi N_{0_2}  \big)^{-n/2}}
\exp \Bigg( - \sum_{k=1}^n \frac{(v_{(2)k} - v_{(1)k} + \mathsf{n}_{(1)k})^2}{2 N_{0_2}}  \Bigg)
;
\end{split}
\end{equation}

Note that also \eqref{eq:pdf_2} can be understood as 
the conditional probability that detector 1 has observed signal $s$
given that detector 2 has observed signal $s$. (See section 2.4 of \cite{Lehmann2005}.)
Then, using Bayes' rule\cite{vanKampen2007} one would also arrive at
\eqref{eq:joint-1}.

The likelihood ratio is calculated from the quotient of
$p(\mathbf{v}_1,s_1,\mathbf{v}_2,s_1)$
with the corresponding $p_0(\mathbf{v}_1,\mathbf{v}_2)$;
where in $p_0$ there is no contribution from any signal
and is given by
\begin{equation}\label{eq:joint-1_0}
\begin{split}
p(\mathbf{v}_1,\mathbf{v}_2)
=&
\frac{1}{\big( 2 \pi N_{0_1}  \big)^{-n/2}}
\exp \Bigg( - \sum_{k=1}^n \frac{(v_{(1)k})^2}{2 N_{0_1}}  \Bigg)
\\
&
\frac{1}{\big( 2 \pi N_{0_2}  \big)^{-n/2}}
\exp \Bigg( - \sum_{k=1}^n \frac{(v_{(2)k} )^2}{2 N_{0_2}}  \Bigg)
.
\end{split}
\end{equation}
This quotient has the form of a product $\mathbf{L}_1 \, \mathbf{L}_2$.

Let us start by calculating the likelihood $\mathbf{L}_1$ considering
just one of these factors, and, as above, using $s_1=s_2$, so that
\begin{equation}
\begin{split}
\mathbf{L}_1( \mathbf{v}_1) =& \frac{p_1(\mathbf{v}_1)}{p_0(\mathbf{v}_1)} \\
=&
\exp \Bigg[ \sum_{k=1}^n 
\frac{2 v_{(1)k}\, v_{(2)k} - v_{(2)k}^2 -2 v_{(1)k}\, \mathsf{n}_{(2)k}
	+ 2 v_{(2)k}\, \mathsf{n}_{(2)k} - \mathsf{n}_{(2)k}^2
}{2 N_{0_1}}   
\Bigg] 
.
\end{split}
\end{equation}

It is convenient to manage the algebra in the following way:
\begin{equation}
\begin{split}
\sum_{k=1}^n &
2 v_{(1)k}\, v_{(2)k} - v_{(2)k}^2 -2 v_{(1)k}\, \mathsf{n}_{(2)k}
+ 2 v_{(2)k}\, \mathsf{n}_{(2)k} - \mathsf{n}_{(2)k}^2
\\
=
\sum_{k=1}^n &
2 v_{(1)k}\, v_{(2)k} 
- v_{(2)k}( v_{(1)k} + \mathsf{n}_{(2)k} - \mathsf{n}_{(1)k} )
-2 v_{(1)k}\, \mathsf{n}_{(2)k}
+ 2 v_{(2)k}\, \mathsf{n}_{(2)k} - \mathsf{n}_{(2)k}^2
\\
=
\sum_{k=1}^n &
2 v_{(1)k}\, v_{(2)k} 
- v_{(2)k} v_{(1)k} - v_{(2)k} \mathsf{n}_{(2)k} + v_{(2)k} \mathsf{n}_{(1)k} \\
&
-2 (s_{(1)k} + \mathsf{n}_{(1)k})\, \mathsf{n}_{(2)k}
+ 2 ( s_{(2)k} + \mathsf{n}_{(2)k})\, \mathsf{n}_{(2)k} - \mathsf{n}_{(2)k}^2 \\
=
\sum_{k=1}^n &
v_{(1)k}\, v_{(2)k} 
- ( s_{(2)k} + \mathsf{n}_{(2)k}) \mathsf{n}_{(2)k} + ( s_{(2)k} + \mathsf{n}_{(2)k}) \mathsf{n}_{(1)k} \\
&
-2 (s_{(1)k} + \mathsf{n}_{(1)k})\, \mathsf{n}_{(2)k}
+ 2 ( s_{(2)k} + \mathsf{n}_{(2)k})\, \mathsf{n}_{(2)k} - \mathsf{n}_{(2)k}^2  \\
=
\sum_{k=1}^n &
v_{(1)k}\, v_{(2)k} 
-  s_{(2)k}  \mathsf{n}_{(2)k} - \mathsf{n}_{(2)k}^2 +  s_{(2)k} \mathsf{n}_{(1)k} + \mathsf{n}_{(2)k} \mathsf{n}_{(1)k} \\
&
-2 s_{(1)k}  \mathsf{n}_{(2)k} - 2  \mathsf{n}_{(1)k}\, \mathsf{n}_{(2)k} 
+ 2  s_{(2)k} \mathsf{n}_{(2)k} + 2 \mathsf{n}_{(2)k}^2 - \mathsf{n}_{(2)k}^2  \\
=
\sum_{k=1}^n &
v_{(1)k}\, v_{(2)k} 
+  s_{(2)k} \mathsf{n}_{(1)k} +   s_{(2)k} \mathsf{n}_{(2)k}  
-2 s_{(1)k}  \mathsf{n}_{(2)k} -   \mathsf{n}_{(1)k}\, \mathsf{n}_{(2)k} 
.
\end{split}
\end{equation}
In this algebraic manipulation we have kept the identities of $s_1$ and $s_2$ to allow
for the situation $s_2 = s_1 + \epsilon$, with $\max|\epsilon| \ll \max|s_1|$.

Since by assumption the noises $\mathsf{n}_1$ and $\mathsf{n}_2$ are considered to have
independent Gaussian behavior; we can take the size of the interval big enough
so that we attain
\begin{equation}\label{key}
\sum_{k=1}^n s_{(2)k} \mathsf{n}_{(1)k} 
\approx \sum_{k=1}^n  s_{(2)k} \mathsf{n}_{(2)k}  
\approx \sum_{k=1}^n s_{(1)k}  \mathsf{n}_{(2)k}
,
\end{equation}
and therefore
\begin{equation}\label{key}
\sum_{k=1}^n s_{(2)k} \mathsf{n}_{(1)k} 
+ \sum_{k=1}^n  s_{(2)k} \mathsf{n}_{(2)k}  
-2 \sum_{k=1}^n s_{(1)k}  \mathsf{n}_{(2)k}
\approx 0
,
\end{equation}
and we also have
\begin{equation}\label{key}
\sum_{k=1}^n \mathsf{n}_{(1)k}\, \mathsf{n}_{(2)k} \approx 0
.
\end{equation}
So that we arrive at
\begin{equation}\label{eq:like-2}
\begin{split}
\mathbf{L}_1( \mathbf{v}_1) =& 
\exp \Bigg[ \sum_{k=1}^n 
\frac{ v_{(1)k}\, v_{(2)k} 
}{2 N_{0_1}}   
\Bigg]
;
\end{split}
\end{equation}
for the likelihood of having the signal $s$ in $v_1$ that is contained in $v_2$.
From the considerations above, we deduce that
the joint likelihood of having the same signal in both detectors is then
\begin{equation}\label{eq:like-2t-1}
\begin{split}
\mathbf{L}_a( \mathbf{v}_1, \mathbf{v}_2) =& 
\exp \Bigg[\frac{1}{2} \Big(  \frac{ 1 }{ N_{0_1}} + \frac{ 1 }{ N_{0_2}} \Big)
\sum_{k=1}^n    v_{(1)k}\, v_{(2)k}
\Bigg]
;
\end{split}
\end{equation}
where in practical situation we evaluate $N_0$ from the sample variance 
so that the final expression is
\begin{equation}\tag{\ref{eq:like-2t}}
\begin{split}
\mathbf{L}( \mathbf{v}_1, \mathbf{v}_2) =& 
\exp \Bigg[\frac{m-1}{2} \Big(  \frac{ 1 }{ \sum_{k=1}^m v_{(1)k}^2 } 
+ \frac{ 1 }{ \sum_{k=1}^m v_{(2)k}^2 } \Big)
\sum_{k=1}^n v_{(1)k}\, v_{(2)k}
\Bigg]
;
\end{split}
\end{equation}
where the width of the window to calculate $\sigma^2$, that is $m$,
is chosen appropriately depending on the nature of the observations $v_k$,
and we are assuming that the means are zero.

\begin{figure}[H]
\centering
\includegraphics[clip,width=0.6\textwidth]{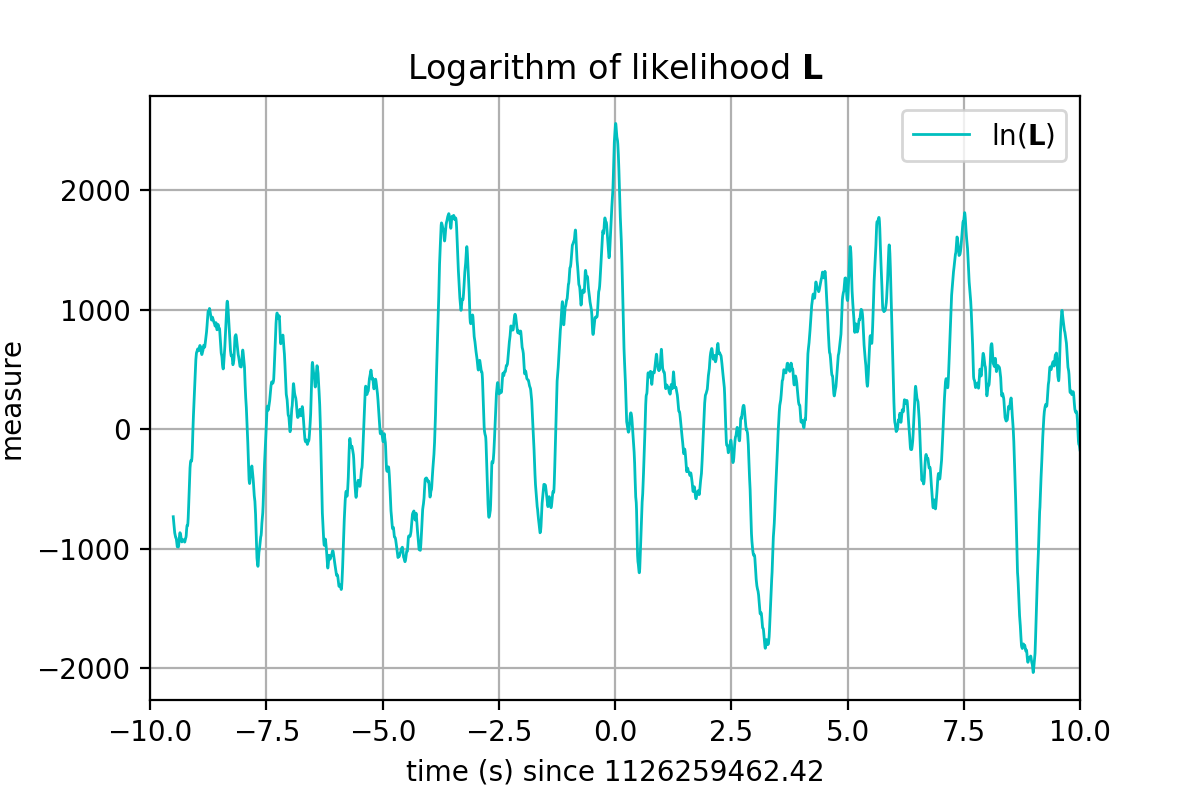}
\caption{Logarithm of the likelihood ratio applied to the strain at Livingston
	and the one at Hanford from -10s to 10s around the time of the event
	with the nominal shift of -0.007s.
}
\label{fig:log-likelihood_007}
\end{figure}

This estimation of the desired measure have some difficulties.
The exponent is huge, when using actual LIGO data,
and it does not emphasize the comparison of the data in both strains.
To understand this in detail, let us see what is the behavior
of the logarithm of this measure,
which we shown in figure \ref{fig:log-likelihood_007};
where we have advanced the time axis by 
by the width of the window.

\subsection{The new measure}

Due to the unpleasant behavior of the likelihood ratio discussed above,
we decide to define a new measure and
so we introduce some changes to likelihood ratio to
strengthen the comparison and to moderate the
amplitude, so that it become useful for the system we have in mind.

To motivate our choice,
let us note that for pure and independent noise one has
\begin{equation}\label{key}
\overline{(\mathsf{n}_1 - \mathsf{n}_2)^2} =
\overline{(\mathsf{n}_1)^2} + \overline{(\mathsf{n}_2)^2} 
- 2 \overline{(\mathsf{n}_1  \mathsf{n}_2)}
=
\overline{(\mathsf{n}_1)^2} + \overline{(\mathsf{n}_2)^2}
= \sigma_1^2 + \sigma_2^2
.
\end{equation}
Also note that when both variances are similar, then
one has that
$\frac{1}{\sigma_1^2} + \frac{1}{\sigma_2^2} $
is approximately
$\frac{4}{\sigma_1 ^2 + \sigma_2^2}$. 
So that instead of $\frac{ 1 }{ N_{0_1}} + \frac{ 1 }{ N_{0_2}}$ 
we can use the estimation
$\frac{4}{   \overline{(v_1 - v_2)^2} }$;
which it has the advantage that
$v_1 -v_2$ would cancel the information of the signal;
and therefore help in accentuating the measure at the
time of coincidence.

In this way we arrive at the measure given by the accentuated likelihood
\begin{equation}\label{eq:Lambda_b}
\begin{split}
\Lambda_a( \mathbf{v}_1, \mathbf{v}_2) =& 
\exp \Bigg[ \Big(  \frac{ 2 (m-1) }{ \sum_{j=1}^m (v_{(1)j} - v_{(2)j})^2 } 
\Big)
\sum_{k=1}^n v_{(1)k}\, v_{(2)k}
\Bigg]
.
\end{split}
\end{equation}

When testing this measure with synthetic noise and artificial signals
it gives huge quantities that are difficult to calculate.
Then, in order to control this too sensitive behavior, we moderate
the measure to be
\begin{equation}\label{eq:Lambda_final_c}
\begin{split}
\Lambda_b( \mathbf{v}_1, \mathbf{v}_2) =& 
\exp \Bigg[ \frac{1}{\sigma^*} \Big(  \frac{  1 }{ \sum_{j=1}^m (v_{(1)j} - v_{(2)j})^2 } 
\Big)
\sum_{k=1}^n v_{(1)k}\, v_{(2)k}
\Bigg]
,
\end{split}
\end{equation}
where $\sigma^*$ is the standard deviation of
$\Big(  \frac{  1 }{ \sum_{j=1}^m (v_{(1)j} - v_{(2)j})^2 } 
\Big)
\sum_{k=1}^n v_{(1)k}\, v_{(2)k}$
in the lapse of time of interest.
Likewise we could have used the factor $\frac{1}{\tilde{\sigma}^2}$, where $\tilde{\sigma}^2$
is the variance of the square root of the exponent in \eqref{eq:Lambda_b}
in the lapse of time of interest.
Note that since we are dividing by the standard deviation,
the factor $2(m-1)$ in the previous expression is immaterial in \eqref{eq:Lambda_final_c}.

Employing the notation of the inner product as in \eqref{eq:inner-app-1},
which makes use of the window $w$,
we arrive at the final expression for the measure given by
\begin{equation}\label{eq:Lambda_final}
\begin{split}
\Lambda( \mathbf{v}_1, \mathbf{v}_2) =& 
\exp \Bigg[ \frac{1}{\sigma^*} 
\frac{ < \mathbf{v}_1 , \mathbf{v}_2 >
}{ 
	< ( \mathbf{v}_1 -\mathbf{v}_2), ( \mathbf{v}_1 -\mathbf{v}_2) > }   
\Bigg]
.
\end{split}
\end{equation}

This measure gives reasonable results with actual LIGO data.


\section{Behavior of the correlation coefficient}\label{sec:apendix-2-correlation}

A natural question is whether the correlation coefficient
between the two set of data coming from both detectors, is enough to have
a well behaved measure for determining if there is a common signal
in the strains.
For this reason we here present the graphs that show the behavior
of the coefficient:
\begin{equation}\tag{\ref{eq:corr-coeff}}
\rho_{\mathbf{v}_1 , \mathbf{v}_2} \equiv  \frac{ < \mathbf{v}_1 , \mathbf{v}_2 >
}{ \sqrt{< \mathbf{v}_1 , \mathbf{v}_1 > < \mathbf{v}_2 , \mathbf{v}_2 > } } 
;
\end{equation}
where we are assuming zero average for both strains.
This measure is also called the sample correlation coefficient\cite{Helstrom75}.
\begin{figure}[H]
\centering
\includegraphics[clip,width=0.45\textwidth]{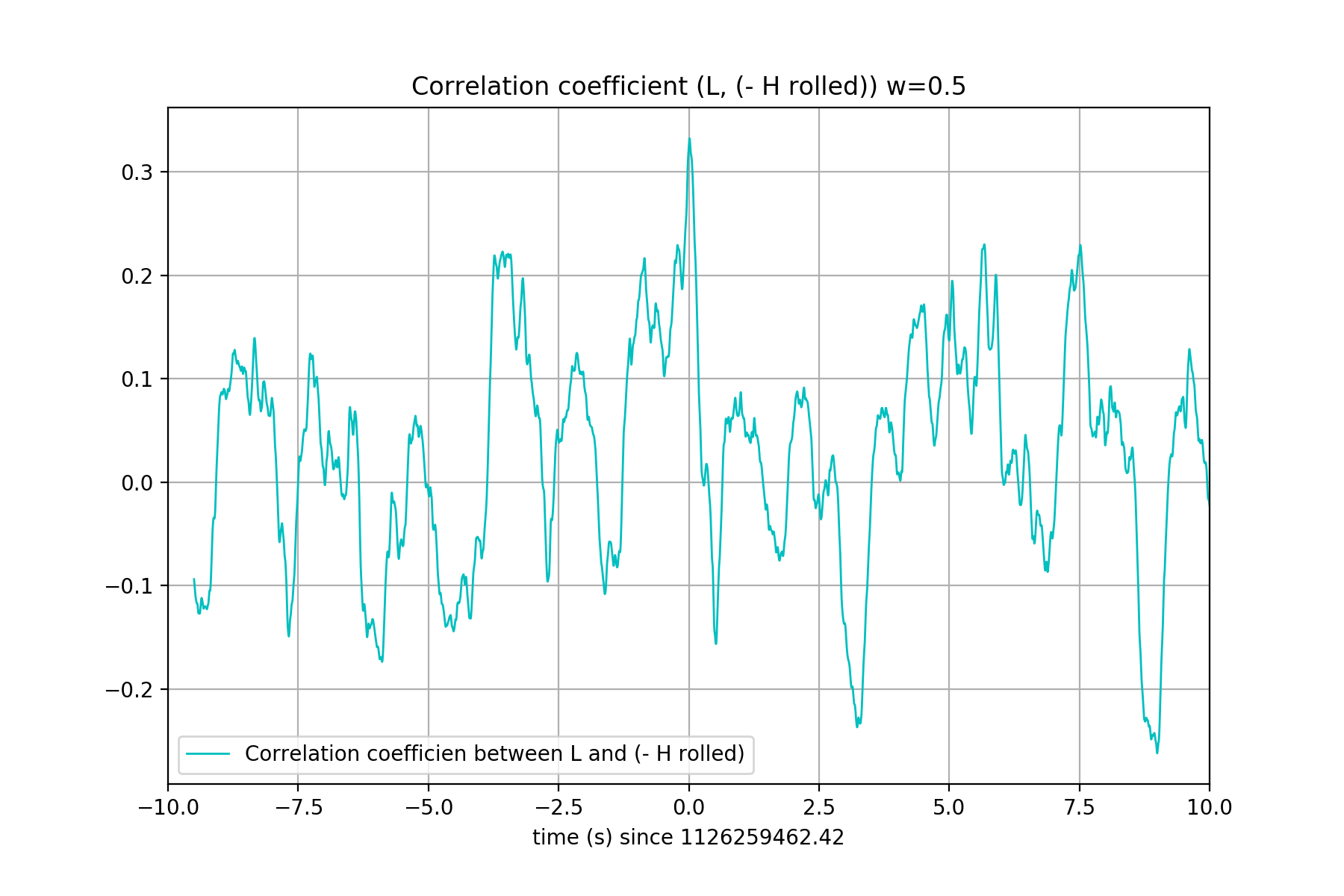}
\includegraphics[clip,width=0.45\textwidth]{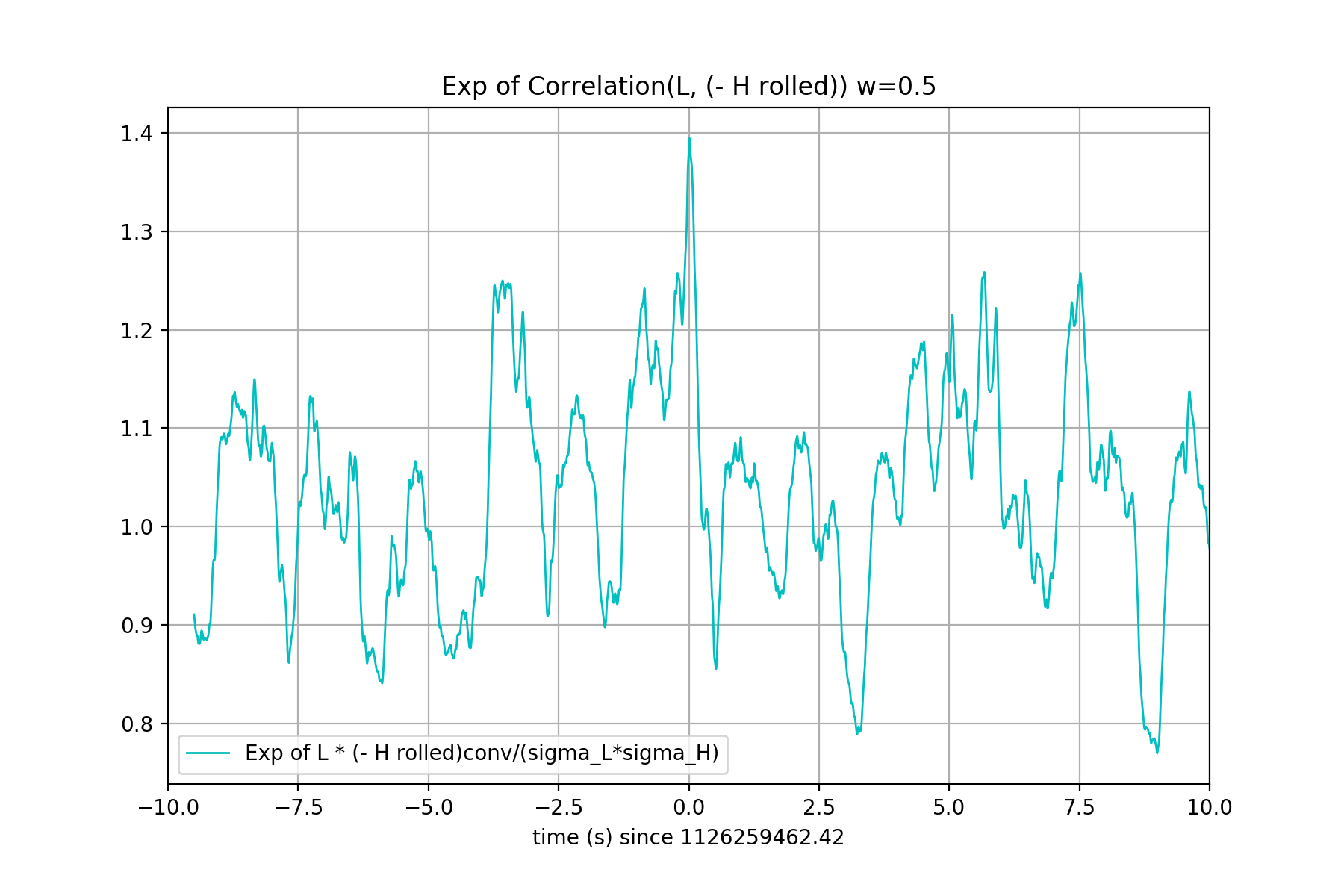}
\caption{On the left the correlation coefficient between the strain at Livingston
	and the one at Hanford from -10s to 10s around the time of the event.
	On the right the exponential of the correlation coefficient
	in the same lapse of time.
}
\label{fig:correlation_007}
\end{figure}
In figure \ref{fig:correlation_007} it is shown the behavior of the correlation coefficient,
where we have made use of the same window employed for the $\Lambda$ measure,
and where we have advanced the time axis
by the width of the window.
We also show the behavior of the exponentiation of the correlation coefficient,
to see if the relation was augmented; but it can be seen that although
there is a local maximum close to the time of the event,
both graphs give information with too much noise.
So, comparing this with the cleaner behavior of our measure $\Lambda$,
as shown in figure \ref{fig:likelihood-1}, we choose $\Lambda$;
which provides a much better tool for analysis.



\end{document}